\documentclass[aps,twocolumn,showpacs,preprintnumbers,amsmath,amssymb,superscriptaddress]{revtex4-1}
\usepackage{amsmath}
\usepackage{bm}
\usepackage{graphicx}
\usepackage{dcolumn}
\usepackage{color}
\usepackage{float}
\usepackage{epstopdf}

\newcommand{\be}{\begin{equation}}
\newcommand{\ee}{\end{equation}}
\newcommand{\bea}{\begin{eqnarray}}
\newcommand{\eea}{\end{eqnarray}}
\newcommand{\bs}{\begin{split}}
\newcommand{\bse}{\begin{subequations}}
\newcommand{\ese}{\end{subequations}}

\begin{document}
\title{Superconductivity in \bm{${\rm Ru_{0.55}Rh_{0.45}P}$} and \bm{${\rm Ru_{0.75}Rh_{0.25}As}$} probed by muon spin relaxation and rotation measurements}

\author{V.~K.~Anand}
\altaffiliation{vivekkranand@gmail.com}
\affiliation{\mbox{Helmholtz-Zentrum Berlin f\"{u}r Materialien und Energie GmbH, Hahn-Meitner Platz 1, D-14109 Berlin, Germany}}
\author{D.~T.~Adroja}
\altaffiliation{devashibhai.adroja@stfc.ac.uk}
\affiliation{ISIS Facility, Rutherford Appleton Laboratory, Chilton, Didcot, Oxon, OX11 0QX, United Kingdom}
\affiliation {Highly Correlated Matter Research Group, Physics Department, University of Johannesburg, P.O. Box 524, Auckland Park 2006, South Africa}
\author{M.~R.~Lees}
\affiliation{Department of Physics, University of Warwick, Coventry CV4 7AL, United Kingdom}
\author{P.~K.~Biswas}
\author{A.~D.~Hillier}
\affiliation{ISIS Facility, Rutherford Appleton Laboratory, Chilton, Didcot, Oxon, OX11 0QX, United Kingdom}
\author{B.~Lake}
\affiliation{\mbox{Helmholtz-Zentrum Berlin f\"{u}r Materialien und Energie GmbH, Hahn-Meitner Platz 1, D-14109 Berlin, Germany}}

\date{\today}

\begin{abstract}
Superconductivity in the pseudo-binary pnictides ${\rm Ru_{0.55}Rh_{0.45}P}$ and ${\rm Ru_{0.75}Rh_{0.25}As}$ is probed by muon spin relaxation and rotation ($\mu$SR) measurements in conjuction with magnetic susceptibility, heat capacity and electrical resistivity measurements. Powder x-ray diffraction confirmed the MnP-type orthorhombic structure (space group $Pnma$) and showed a nearly single phase nature with small impurity phase(s) of about 5\% for both the samples. The occurence of bulk superconductivity is confirmed with $T_{\rm c} = 3.7$~K for ${\rm Ru_{0.55}Rh_{0.45}P}$ and $T_{\rm c} =1.6$~K for ${\rm Ru_{0.75}Rh_{0.25}As}$. The superconducting state electronic heat capacity data reveal weak-coupling single-band isotropic $s$-wave gap BCS superconductivity. Various normal and superconducting state parameters are determined which reveal a weak-coupling electron-phonon driven type-II dirty-limit superconductivity for both the compounds. The upper critical field shows a linear temperature dependence down to the lowest measured temperatures which is quite unusual for a single-band superconductor. The $\mu$SR data confirm the conventional type-II behavior, and show evidence for a single-band $s$-wave singlet pairing superconductivity with a preserved time reversal symmetry for both the compounds.
\end{abstract}

\maketitle

\section{\label{Intro} Introduction}

The discovery of superconductivity in FeAs-based compounds stimulated great interest in pnictide materials \cite{Johnston2010,Stewart2011}. Recently the pseudo-binary pnictides Ru$_{1-x}$Rh$_xPn$ ($Pn$ = P, As) which are free of iron were reported to show superconductivity \cite{Hirai2012}. Interestingly, the parent compounds RuP and RuAs are nonsuperconducting and nonmagnetic, implying that the superconductivity in these pseudo-binary pnictides is accessed through a nonmagnetic critical point. The nonmagnetic route to superconductivity in these pseudo-binary pnictides is distinct from that of iron arsenides, where superconductivity occurs upon suppressing the ordered Fe moment, making them very interesting for further investigations that should be helpful in understanding the physics of superconductivity in pnictides and ascertain the role of Fe moment in iron arsenide superconductors.

\begin{figure}
\includegraphics[width=4cm, keepaspectratio]{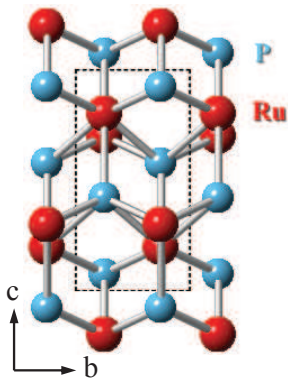}
\caption{MnP-type orthorhombic structure (space group {\it Pnma}) of RuP as viewed along the {\it a} axis.}
\label{fig:Struct}
\end{figure}

Both RuP and RuAs crystallize with a MnP-type orthorhombic structure (space group {\it Pnma}) which consists of face-sharing chains of RuPn$_6$ octahedra along the {\it a}-axis and a distorted triangular lattice of Ru within the {\it bc} plane \cite{Hirai2012}. The crystal structure is illustrated in Fig.~\ref{fig:Struct}. Both RuP and RuAs have nonmagnetic and nonsuperconducting ground states, though, they undergo a metal to insulator transition below 270~K (RuP) and 200~K (RuAs)  \cite{Hirai2012}. Furthermore, they also exhibit evidence for the pseudogap formation associated with a structural phase transition at 330~K for RuP and 280~K for RuAs \cite{Hirai2012}. The partial substitution of Ru by Rh suppresses both pseudogap formation and metal-insulator transition, leading to the emergence of superconductivity with a maximum $T_{\rm c}$ of 3.7~K for ${\rm Ru_{0.55}Rh_{0.45}P}$ and 1.8~K for ${\rm Ru_{0.75}Rh_{0.25}As}$ \cite{Hirai2012}.

In order to characterize the superconducting properties of ${\rm Ru_{0.55}Rh_{0.45}P}$ and ${\rm Ru_{0.75}Rh_{0.25}As}$ in detail we have investigated the physical properties of these two pseudo-binaries by means of various complementary tools. Here we report our results on the superconducting and normal state properties of  ${\rm Ru_{0.55}Rh_{0.45}P}$ and ${\rm Ru_{0.75}Rh_{0.25}As}$ based on magnetic susceptibility $ \chi(T)$,  isothermal magnetization $M(H)$, heat capacity $C_{\rm p}(T,H)$, electrical resistivity $\rho(T,H)$ and muon spin relaxation and rotation ($\mu$SR) measurements. Our $M(T)$, $C_{\rm p}(T)$ and $\rho(T)$ data confirm the bulk superconductivity with $T_{\rm c} = 3.7$~K for ${\rm Ru_{0.55}Rh_{0.45}P}$ and $T_{\rm c} = 1.6$~K for ${\rm Ru_{0.75}Rh_{0.25}As}$. The superconducting state electronic heat capacity of both ${\rm Ru_{0.55}Rh_{0.45}P}$ and ${\rm Ru_{0.75}Rh_{0.25}As}$ can be described by the conventional single-band weak coupling BCS model of superconductivity. The superconducting state parameters characterize them as weakly coupled electron-phonon driven type-II superconductors in the dirty-limit. Our $\mu$SR data further confirm the type-II superconductivity with a single-band $s$-wave singlet pairing and preserved time reversal symmetry in both the compounds. 

\section{\label{ExpDetails} Experimental Details}

Polycrystalline samples of ${\rm Ru_{0.55}Rh_{0.45}P}$ and ${\rm Ru_{0.75}Rh_{0.25}As}$ were prepared by the solid state reaction method at the Core Lab for Quantum Materials, Helmholtz-Zentrum Berlin. Stoichiometric amounts of high purity elements (Ru:  99.9\%, Rh: 99.99\%, P: 99.95\%, As: 99.999\%) in powder form were mixed and ground, pelletized and sealed in quartz tubes, and then sintered at 1100 $^{\circ}$C (${\rm Ru_{0.55}Rh_{0.45}P}$) and 1000 $^{\circ}$C (${\rm Ru_{0.75}Rh_{0.25}As}$) for 60~h\@. The samples were reground, pelletized, sealed in quartz tubes, and sintered for 80~h at the same temperatures used for the first heat treatment. The samples quality and crystallographic information were checked by powder x-ray diffraction (XRD) using Cu K$_{\alpha}$ radiation. 

The room temperature powder XRD patterns revealed a nearly single phase nature of both the samples with small impurity phase(s) of about 5\%. It is evident from the magnetic susceptibility and the zero-field $\mu$SR measurements that these impurities are principally nonmagnetic. The Rietveld refinement with MnP-type orthorhombic structure (space group {\it Pnma}) yielded lattice parameters $a = 5.4230(4) $~{\AA},  $b = 3.3891(3)$~{\AA} and $c = 5.9255(4)$~{\AA} for ${\rm Ru_{0.55}Rh_{0.45}P}$ and $a = 5.6322(3) $~{\AA},  $b = 3.4730(2)$~{\AA} and $c = 6.2065(3)$~{\AA} for ${\rm Ru_{0.75}Rh_{0.25}As}$. 

The magnetic susceptibility and isothermal magnetization were measured using a Quantum Design Magnetic Property Measurement System (MPMS) SQUID magnetometer. The heat capacity measurements were performed by the relaxation method using a Quantum Design Physical Property Measurement System (PPMS). The electrical resistivity measurements were performed by a standard four-probe ac technique using the PPMS. Temperatures down to 0.35~K were attained by a $^3$He insert in the PPMS. For magnetic properties we use Gaussian cgs units, where tesla ($1~{\rm T} = 10^4$~Oe) is a unit of convenience for magnetic field $H$.

The muon spectroscopy measurements were carried out using the MuSR spectrometer at the ISIS facility of the Rutherford Appleton Laboratory, United Kingdom, with the detectors in both longitudinal and transverse configurations. A high purity silver (99.999\%) plate, which only gives a non-relaxing background signal, was used to mount the sample. The powdered samples were mounted on the silver plates using diluted General Electric (GE) varnish and then covered with thin silver foils. Temperatures down to 50~mK were achieved by cooling the sample in a dilution refrigerator. Correction coils were used to cancel the stray fields at the sample position to within 1 $\mu$T.

\section{\label{Sec:RuRhP} Superconductivity in \bm{${\rm R\lowercase{u}_{0.55}R\lowercase{h}_{0.45}P}$}}

\subsection{\label{Sec:RuRhP_MT} Magnetic Susceptibility and Magnetization}

\begin{figure}
\includegraphics[width=8cm]{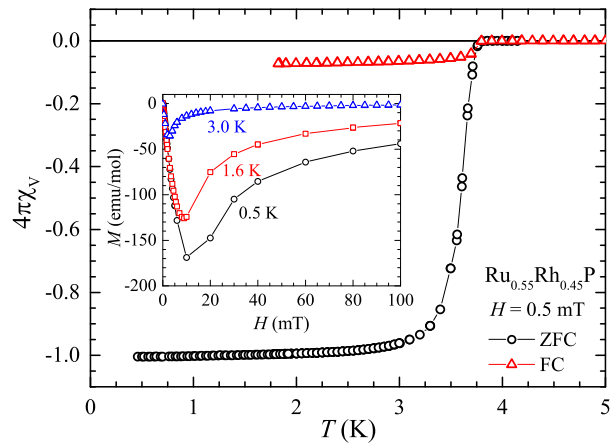} 
\caption{(Color online) Zero field cooled (ZFC) and field cooled (FC) magnetic susceptibility $\chi$ of  ${\rm Ru_{0.55}Rh_{0.45}P}$ as a function of temperature~$T$ for 0.46~K~$\leq T \leq$~5~K measured in applied magnetic field $H=0.5$~mT. Inset: Isothermal magnetization $M(H)$ at $T=0.5$, 1.6 and 3.0~K\@.}
\label{fig:RuRhP_chi}
\end{figure}

The zero field cooled (ZFC) and field cooled (FC) $\chi(T)$ data for ${\rm Ru_{0.55}Rh_{0.45}P}$ measured in $H=0.5$~mT are shown in Fig.~\ref{fig:RuRhP_chi}. A clear superconducting transition near 3.7~K is evident from both ZFC and FC $\chi(T)$. The large Meissner signal for the ZFC $\chi$ corresponds to almost $100$\% superconducting phase fraction, revealing bulk superconductivity in ${\rm Ru_{0.55}Rh_{0.45}P}$. A large Meissner signal is also seen in the isothermal $M(H)$ data at $T=0.5$~K (see inset of Fig.~\ref{fig:RuRhP_chi}). It is seen that the $M$ is initially linear in $H$ and deviates from this linear behavior as $H$ increases further. This deviation from the linearity of $M(H)$ at low-$H$ marks the lower critical field $H_{c1}$ ($\sim 5.6$~mT at 0.5~K) which as expected decreases as the temperature approaches to $T_{\rm c}$, e.g., at 1.6~K and 3.0~K (see inset of Fig.~\ref{fig:RuRhP_chi}).

\begin{figure}
\includegraphics[width=8cm]{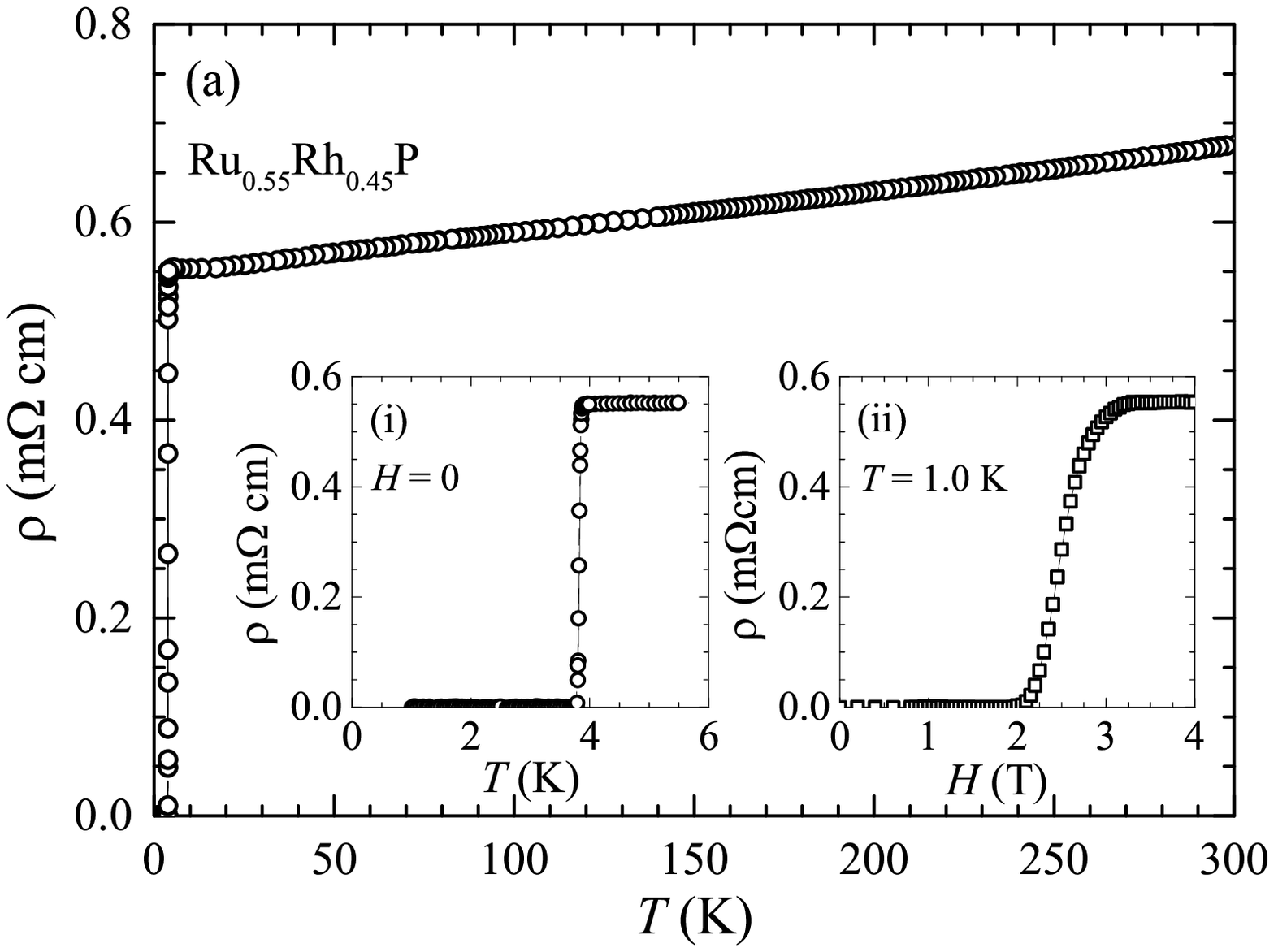}\vspace{0.1cm}
\includegraphics[width=8cm]{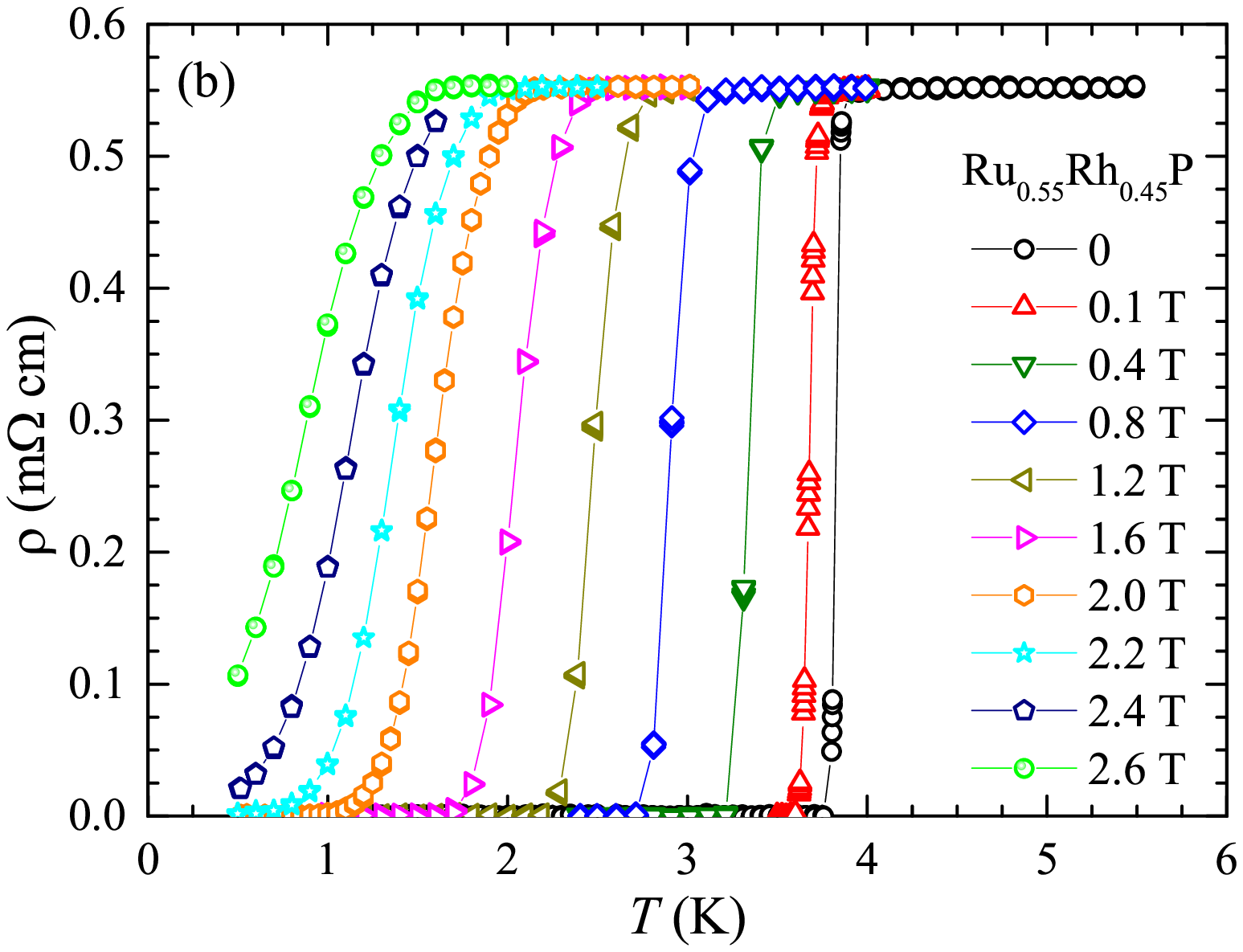} 
\caption{(Color online) (a) Electrical resistivity $\rho$ of ${\rm Ru_{0.55}Rh_{0.45}P}$ as a function of temperature~$T$ for 1~K~$\leq T \leq$~300~K measured in applied magnetic field $H=0$. Insets: (i) Expanded plot of $\rho(T)$ showing the superconducting transition, and  (ii) $\rho(H)$ at 1~K\@. (b) Low-$T$ $\rho(T)$ at different $H$ for 0.45~K~$\leq T \leq$~5.5~K.}
\label{fig:RuRhP_rho}
\end{figure}

\subsection{\label{Sec:RuRhP_Rho} Electrical Resistivity}

The $\rho(T)$ data for ${\rm Ru_{0.55}Rh_{0.45}P}$ measured at various fields are shown in Fig.~\ref{fig:RuRhP_rho}. A clear superconducting transition is seen in $\rho(T)$. In the normal state, the $\rho(T)$ data reveal a metallic character, i.e. the $\rho$ decreases with decreasing $T$, reaching a value of 0.55~m$\Omega$\,cm at 5~K giving a residual resistivity ratio of 1.2. The onset of superconductivity occurs at $T_{\rm c}^{\rm  onset} \approx 3.9$~K and the zero resistance state is reached at $T_{\rm c\, 0} \approx 3.8$~K\@ [see inset (i) of Fig.~\ref{fig:RuRhP_rho}(a)]. The effect of magnetic field on $T_{\rm c}$ is clear from the $\rho(T)$ measured in different $H$ [Fig.~\ref{fig:RuRhP_rho}(b)], the $T_{\rm c}$ decreases with increasing $H$. The $\rho(H)$ data indicate that a field of about 3.2~T is required to completely destroy  the superconductivity [see inset (ii) of Fig.~\ref{fig:RuRhP_rho}(a)].

\subsection{\label{Sec:RuRhP_HC} Heat Capacity}

\begin{figure}
\includegraphics[width=8cm]{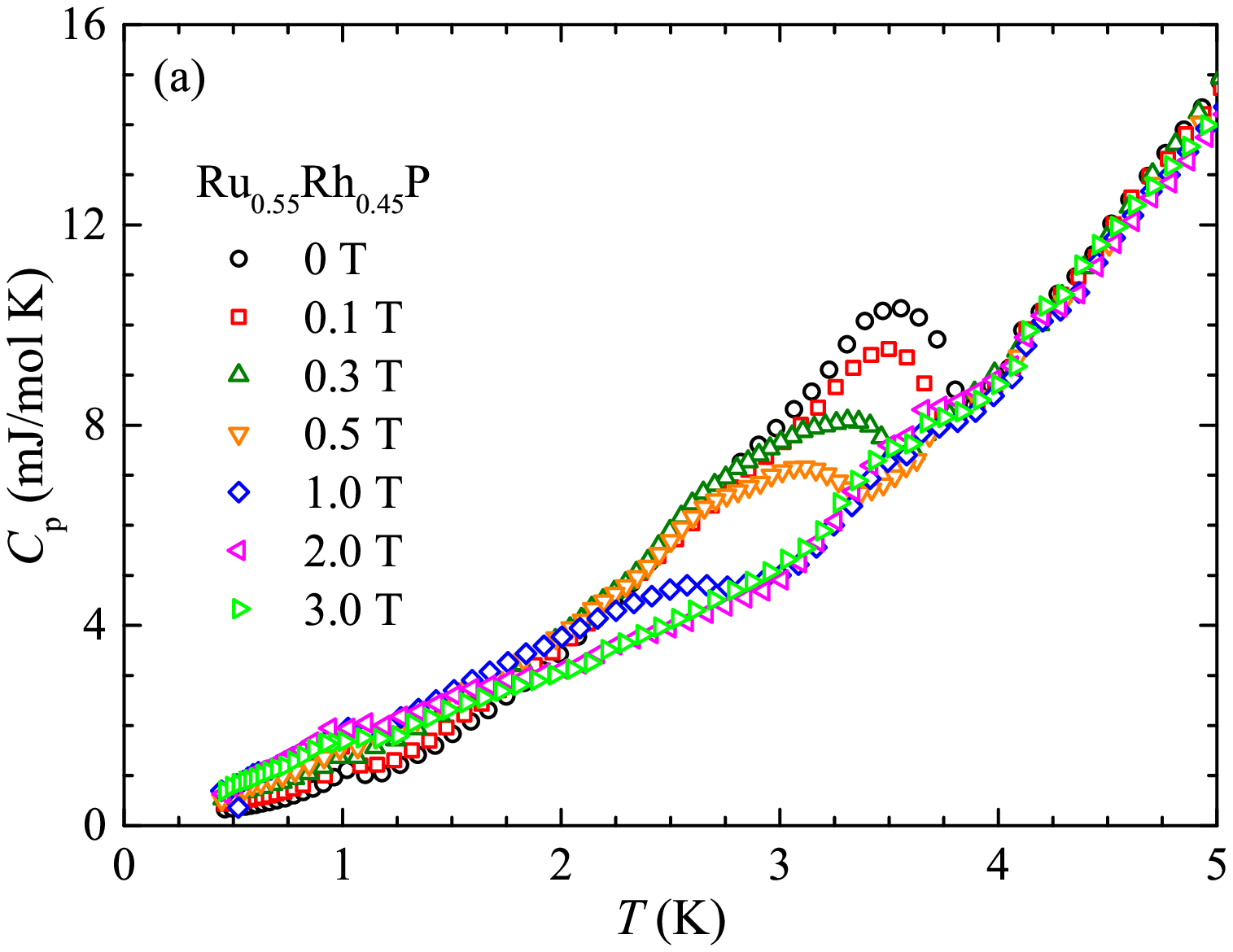}
\includegraphics[width=8cm]{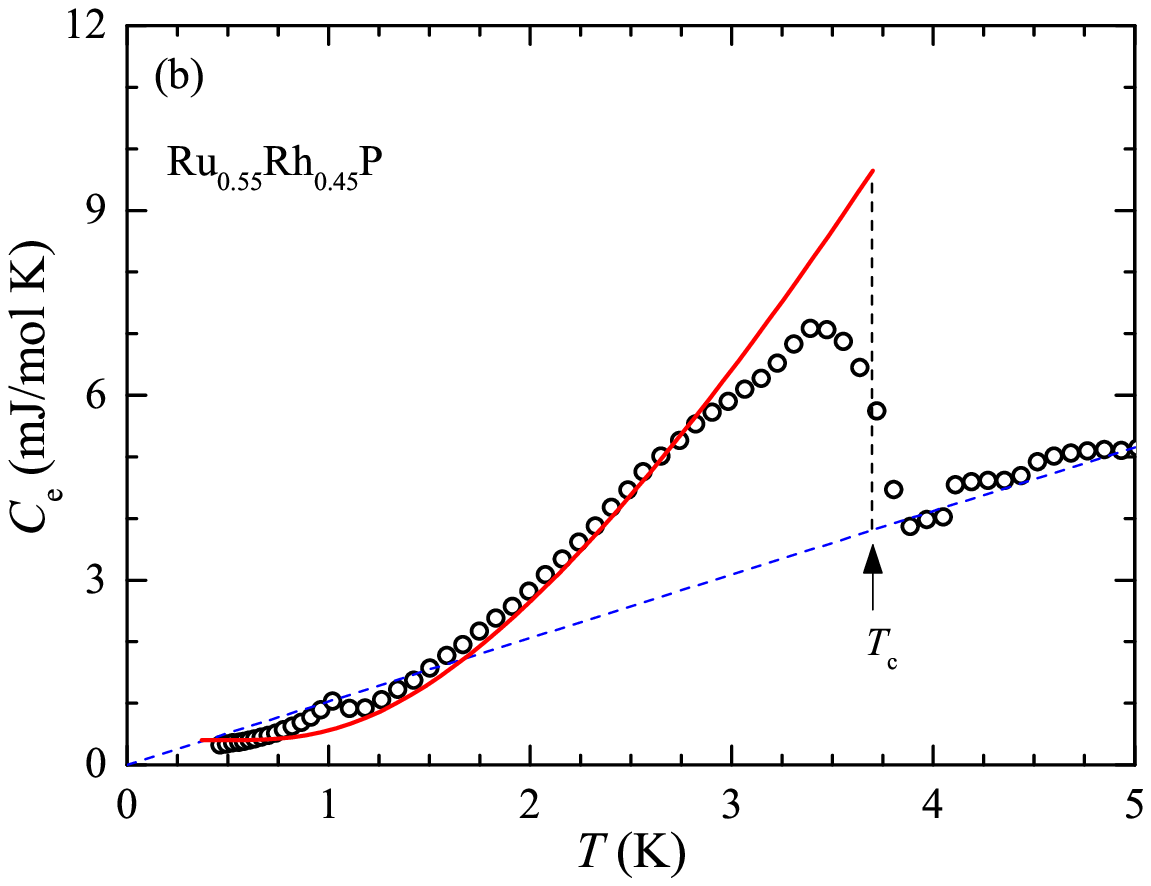}
\caption{(Color online) (a) Heat capacity $C_{\rm p}$ of ${\rm Ru_{0.55}Rh_{0.45}P}$ as a function of temperature $T$ for 0.45~K~$\leq T \leq$~5~K measured in different indicated applied magnetic fields. (b) Electronic contribution $C_{\rm e}(T)$ to zero field heat capacity. The solid red curve is the theoretical prediction for single-band fully gapped ($\Delta(0)/k_{\rm B}T_{\rm c} = 1.764$) BCS superconductivity. The theoretical curve is shifted up by 0.40~mJ/mol~K which accounts for nonsuperconducting contribution to $C_{\rm e}$. The dashed blue line depicts the $\gamma_{\rm n} T$.}
\label{fig:RuRhP_HC}
\end{figure}

The $C_{\rm p}(T)$ data for ${\rm Ru_{0.55}Rh_{0.45}P}$ measured at various fields  are shown in Fig.~\ref{fig:RuRhP_HC}. An anomaly related to the superconducting transition is clearly seen from the $C_{\rm p}(T)$ data, $T_{\rm c}^{\rm onset} = 3.86$~K at $H=0$. Using the entropy-conserving construction [as shown in Fig.~\ref{fig:RuRhP_HC}(b)] we define $T_{\rm c} = 3.70(5)$~K\@. The application of magnetic field suppresses the $T_{\rm c}$, and at $H=3.0$~T the anomaly related to superconductivity is suppressed to a temperature below 0.46~K\@ [see Fig.~\ref{fig:RuRhP_HC}(a)]. We also see an anomaly near 1~K whose origin is not clear and we attribute it to the presence of unidentified impurity in the sample. The absence of any corresponding anomaly in the magnetic susceptibility data or the muon spectroscopy data presented below, supports the view that the bulk of any impurity in the sample is nonmagnetic.  A secondary superconducting phase with a different Rh concentration seems very likely to be the source of this 1 K anomaly in $C_{\rm p}(T)$.

\begin{figure*}
\includegraphics[width=\textwidth, keepaspectratio]{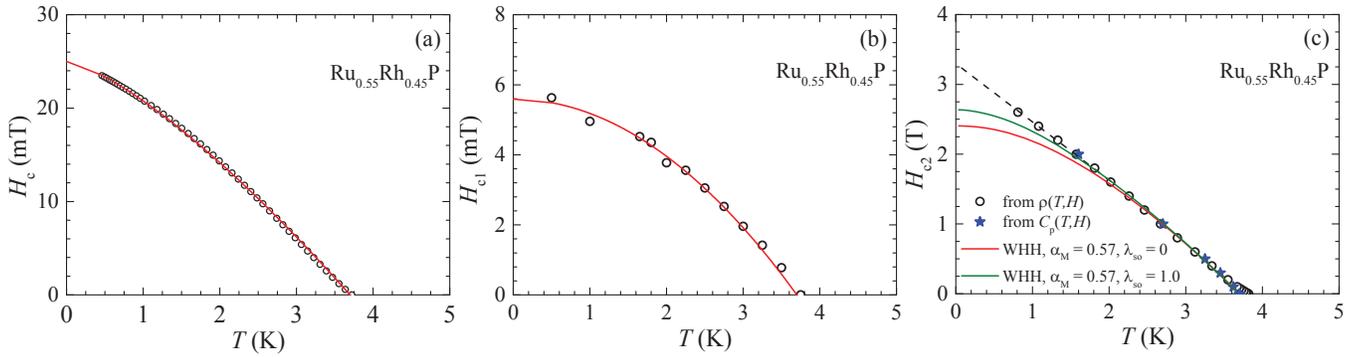}
\caption{(Color online) (a) Thermodynamic critical field $H_{c}$ of ${\rm Ru_{0.55}Rh_{0.45}P}$ as a function of temperature $T$  obtained from the experimental electronic heat capacity $C_{\rm e}(T)$ data. (b) Lower critical field $H_{c1}(T)$ obtained from $M(H)$ data, and (c) Upper critical field $H_{c2}(T)$ obtained from $C_{\rm p}(T,H)$ and $\rho(T,H)$ data. The solid curves represent the fits as discussed in the text. The dashed line in (c) shows a linear behavior.}
\label{fig:RuRhP_HCrit}
\end{figure*}

The low-$T$ $C_{\rm p}(T)$ data above $T_{\rm c}$ are well described by $C_{\rm p}(T) = \gamma_{\rm n} T + \beta T^3$, allowing us to estimate the normal state Sommerfeld coefficient $ \gamma_{\rm n} = 1.03(4)$~mJ/mol\,K$^2$. The coefficient $\beta$ is found to  be 0.078~mJ/mol\,K$^4$ which gives an estimate of Debye temperature $\Theta_{\rm D} = (12 \pi^{4} R n/5 \beta )^{1/3}$  = 368(5)~K, where $R$ is the molar gas constant, and $n=2$ the number of atoms per formula units \cite{Kittel2005}. We estimate the density of states at the Fermi level ${\cal D}(E_{\rm F})$ according to the relation $\gamma_{\rm n} = ({\pi^2 k_{\rm B}^2}/{3}) {\cal D}(E_{\rm F})$, yielding ${\cal D}(E_{\rm F}) = 0.44(1) $ states/eV f.u. for both spin directions. The bare band-structure density of states ${\cal D}_{\rm band}(E_{\rm F})$ can be found using the relation ${\cal D}(E_{\rm F}) = {\cal D}_{\rm band}(E_{\rm F})(1 + \lambda_{\rm e-ph})$ \cite{Grimvall1976}. The electron-phonon coupling constant $\lambda_{\rm {e-ph}}$ can be determined using McMillan's relation \cite{McMillan1968}
\begin{equation}
\lambda_{\rm {e-ph}}= \frac {1.04+\mu^{\ast} \ln(\Theta_{\rm D}/1.45\,T_{\rm c})} {(1-0.62\mu^{\ast})\ln(\Theta_{\rm D}/1.45\,T_{\rm c}) - 1.04}.
\label{eq:lambda}
\end{equation}
Accordingly, for $\mu^{\ast} = 0.13$, and using the values of $T_{\rm c} = 3.7$~K and $\Theta_{\rm D}= 368$~K, we obtain $\lambda_{\rm {e-ph}} = 0.56$. The small value of $\lambda_{\rm {e-ph}}$ reflects a weak-coupling superconductivity in ${\rm Ru_{0.55}Rh_{0.45}P}$. Using $\lambda_{\rm {e-ph}} = 0.56$, we get ${\cal D}_{\rm band}(E_{\rm F}) = 0.28$ states/eV f.u. for both spin directions. The effective quasiparticle mass $m^\ast = m^\ast_{\rm band}(1 + \lambda_{\rm e-ph})$ is estimated to be $m^\ast =1.56 m_{\rm e}$. The Fermi velocity $v_{\rm F}$ estimated using the relation \cite{Kittel2005} $v_{\rm F} = (\pi^2 \hbar^3/{m^{\ast}}^2 V_{\rm f.u.}) {\cal D}(E_{\rm F}) = 5.74 \times 10^7~{\rm cm/s}$, where $V_{\rm f.u.}$ is the volume per formula unit. The mean free path given by \cite{Anand2013a} $\ell = (3\pi^2 \hbar^3)/(e^2 {m^\ast}^2 v_{\rm F}^2 \rho_0) = 0.37~{\rm nm}$. This value of $\ell$ is close to the lattice parameter $b$.

\subsection{\label{Sec:RuRhP_SC_Prop} Superconducting state properties}

In order to estimate the superconducting parameters we separate out the electronic contribution to the heat capacity $C_{\rm e}(T)$ by subtracting off the lattice contribution from the measured $C_{\rm p}(T)$, i.e. $C_{\rm e}(T) = C_{\rm p}(T) - \beta T^3 $. The $C_{\rm e}(T)$  estimated for ${\rm Ru_{0.55}Rh_{0.45}P}$ is shown in Fig.~\ref{fig:RuRhP_HC}(b). The $C_{\rm e}(T)$ shows superconducting transition more clearly, reflecting the bulk nature of superconductivity.  A jump of $\Delta C_{\rm e} = 5.40(5)$~mJ/mol\,K at $T_{\rm c}$ is obtained corresponding to the entropy-conserving construction shown by the vertical dotted line at $T_{\rm c}$ in Fig.~\ref{fig:RuRhP_HC}(b). Accordingly we obtain the parameter $\Delta C_{\rm e}/ \gamma_{\rm n} T_{\rm c} = 1.42(1)$ for $T_{\rm c} = 3.7$~K and  $\gamma_{\rm n} = 1.03(4)$~mJ/mol\,K$^2$, which is in very good agreement with the BCS value of 1.426 in the weak-coupling limit \cite{Tinkham1996}.  

We analyze $C_{\rm e}(T)$ data within the framework of single-band fully-gapped BCS model of superconductivity which is also supported by our $\mu$SR data (discussed latter). The theoretical prediction for the fully gapped, $\Delta(0)/k_{\rm B}T_{\rm c} =1.764$ [where $\Delta(0)$ is the superconducting gap at $T=0$], BCS superconductivity is shown in Fig.~\ref{fig:RuRhP_HC}(b). A reasonable agreement between the experimental data and the theoretical prediction can be seen from Fig.~\ref{fig:RuRhP_HC}(b). In order to compare the experimental data and theoretical prediction, the theoretical curve has been shifted by 0.40~mJ/mol\,K which can be attributed to the presence of small nonsuperconducting impurity phase(s).

We estimate the thermodynamic critical field $H_{\rm c}(T)$ using the zero-field $C_{\rm e}(T)$ data. The $H_{\rm c}$ is related to the entropy difference between the normal $S_{\rm en}$ and superconducting $S_{\rm es}$ states \cite{Tinkham1996, DeGennes1966}, $H_{\rm c}^{2}(T) = 8\pi\int_{T}^{T_{\rm c}}[S_{\rm en}(T^\prime)-S_{\rm es}(T^\prime)] dT^\prime$. The electronic entropies can be estimated by integrating the electronic heat capacity, i.e. $S_{\rm e}(T^{\prime}) = \int_0^{T^\prime}[C_{\rm e}(T^{\prime\prime})/T^{\prime\prime})] dT^{\prime\prime}$. The $H_{\rm c}(T)$ obtained this way is shown in Fig.~\ref{fig:RuRhP_HCrit}(a). The $H_{\rm c}(T)$ data follow the behavior $H_{\rm c}(T) = H_{\rm c}(0) [1-(T/T_{\rm c})^p]$, however with $p = 1.36(1)$ which is much lower than 2. The fit of $H_{\rm c}(T)$ data shown by solid red curve in Fig.~\ref{fig:RuRhP_HCrit}(a) yields $H_{\rm c}(0) = 25.0(1)$~mT. 

The $T$ dependence of the lower critical field $H_{\rm c1}$ determined from the $M(H)$ isotherms collected at various $T$ is shown in Fig.~\ref{fig:RuRhP_HCrit}(b). The $H_{\rm c1}(T)$ data are well described by the conventional behavior $H_{\rm c1}(T) = H_{\rm c1}(0) [1-(T/T_{\rm c})^p]$, with $p = 2$, the fit is shown by the solid red curve in Fig.~\ref{fig:RuRhP_HCrit}(b). Accordingly we obtain $H_{\rm c1}(0) = 5.6(1)$~mT. This value of $H_{\rm c1}(0)$ is much lower than the $H_{\rm c}(0)= 25.0(1)$~mT obtained above, indicating a type-II superconductivity in ${\rm Ru_{0.55}Rh_{0.45}P}$.

The $T$ dependence of the upper critical field $H_{\rm c2}$ determined from the $C_{\rm p}(T,H)$ and $\rho(T,H)$ data is shown in Fig.~\ref{fig:RuRhP_HCrit}(c). The much larger value of $H_{\rm c2} (T$$\rightarrow$$0)$ compared to $H_{\rm c1}(0)$  and $H_{\rm c}(0)$ further confirms the type II superconductivity in ${\rm Ru_{0.55}Rh_{0.45}P}$. The initial slope of $H_{\rm c2} (T)$ is found to be $dH_{\rm c2}(T)/dT|_{T = T\rm c} = -1.08(2)$~T/K. The orbital critical field  $H_{\rm c2}^{\rm Orb} (0)$ estimated according to \cite{Hefland1966,WHH1966} $H_{\rm c2}^{\rm Orb} (0) = -A\, T_{\rm c}\,dH_{\rm c2}(T)/dT|_{T = T\rm c}$ is 2.92(5)~T in the clean limit ($A = 0.73$) and 2.76(5)~T in the dirty limit ($A = 0.69$). The Pauli-limiting upper critical field $H_{\rm P}(0) = 1.86\,T_{\rm c}$ \cite{Clogston1962, Chandrasekhar1962}, accordingly we obtain  $H_{\rm P}(0) = 6.88$~T\@. The Maki parameter $\alpha_{\rm M}= \sqrt{2}\, H_{\rm c2}^{\rm Orb}(0)/H_{\rm P}(0) = 0.57$ \cite{Maki1966} using the dirty limit value of $H_{\rm c2}^{\rm Orb} (0)$. The small value of $\alpha_{\rm M}$ suggests that the orbital pair breaking is important in determining the $H_{\rm c2}$.

\begin{table}
\caption{\label{tab:SCParams} Measured and derived superconducting and relevant normal state parameters for ${\rm Ru_{0.55}Rh_{0.45}P}$ and  ${\rm Ru_{0.55}Rh_{0.45}As}$.  }
\begin{ruledtabular}
\begin{tabular}{lcc}
			 							& ${\rm Ru_{0.55}Rh_{0.45}P}$ &  ${\rm Ru_{0.55}Rh_{0.45}As}$\\
\hline
$T_{\rm c}$ (K)                                         				& 3.70(5)   & 1.60(4) \\
$\gamma_{\rm n}$ (mJ/mol\,K$^{2}$)                      		& 1.03(4)  & 3.79(6)\\
${\cal D}(E_{\rm F})$ (states/eV\,f.u.)						& 0.44(1)	& 1.61(2)	\\
$\Theta_{\rm D}$ (K)									      	& 368(5) & 284(2) \\
$\lambda_{\rm e-ph}$                                    				& 0.56  & 0.49  \\
$\Delta C_{\rm e}$ (mJ/mol\,K)                          			& 5.40(5) &  8.14(8)\\
$\Delta C_{\rm e}/\gamma_{\rm n} T_{\rm c}$             		& 1.42(1)  & 1.42(2)\\
$\Delta(0)/k_{\rm B}T_{\rm c}$~(K) from $\mu$SR	    		& 1.78(3)  & 1.81(6) \\
$\alpha_{\rm M}$                                        				&  0.57  & 0.90\\
$H_{\rm c}(T=0)$ (mT)                                   				&  25.0(1) & 16.6(2) \\
$H_{\rm P}$ (T) 									    		&  6.88 & 2.98\\
$H_{\rm c1}(T=0)$ (mT)                                  			&  5.6(1)  & 5.4(1) \\
$H_{\rm c2}^{\rm Orb}(T=0)$ (T)		         			& 2.76 (5) & 1.90(4) \\
$H_{\rm c2}(T=0)$ (T)                                  				& 3.30(2) & 2.60(1) \\
$\kappa_{\rm GL}$                                       				&  93 & 111\\
$\xi_{\rm GL}(T=0)$ (nm)                                         				&  10  & 11 \\
$\xi_{\rm BCS}(T=0)$ (nm)                                            					&  214   & 1792  \\
$\ell~(m^\ast=1.51m_{\rm e})$ (nm)                          		&  0.37--0.42   & 0.023--0.051 \\
$\lambda_{\rm eff}^{\rm calc}(0)$~(nm)    				&  933  & 1247\\
$\lambda_{\rm eff}^{\rm obs}(0)$ (nm) from $\mu$SR         &  309(3) & 487(4)  \\
\end{tabular}
\end{ruledtabular}
\end{table}

It is seen that the $H_{\rm c2}$ shows a linear $T$ dependence without showing any saturation tendency at low temperatures. This linear behavior of $H_{\rm c2}(T)$ is quite distinct from the behavior of isotropic, single-band BCS superconductors for which $H_{\rm c2}(T)$ exhibits a linear temperature dependence only close to $T_{\rm c}$ and saturates at low temperatures with a downward curvature. As such the $H_{\rm c2}(T)$ could not be described by the Werthamer, Helfand, and Hohenberg (WHH) model for an isotropic superconductor in the dirty limit \cite{Hefland1966, WHH1966}. The WHH model predicted $H_{\rm c2}(T)$ for $\alpha_{\rm M} = 0.57$ and $\lambda_{\rm so} = 0$ as well as $\lambda_{\rm so} = 1.0$ are shown in Fig.~\ref{fig:RuRhP_HCrit}(c). The departure from the WHH model is quite clear at low-$T$. Therefore the upper critical field is estimated by a linear extrapolation of $H_{\rm c2}(T)$, which yields $H_{\rm c2}(0) = 3.30(2)$~T\@. 

The Ginzburg-Landau parameter $\kappa_{\rm GL}= H_{\rm c2}(0)/\sqrt{2}\,H_{\rm c}(0) \approx 93 \gg 1/\sqrt{2}$ for $H_{\rm c2}(0) = 3.30$~T and $H_{\rm c}(0) = 25.0$~mT clearly classifies ${\rm Ru_{0.55}Rh_{0.45}P}$ as a type-II superconductor.  The Ginzburg-Landau coherence length $\xi_{\rm GL}(0)$ can be estimated from \cite{Tinkham1996, DeGennes1966} $H_{\rm c2}(0) = \Phi_0/2\pi \xi_{\rm GL}(0)^2 $, where the flux quantum $\Phi_0 = 2.07 \times 10^{-7}$~G\,cm$^2$. Accordingly, for $H_{\rm c2(0)} = 3.30$~T we get $\xi_{\rm GL}(0) = 10~{\rm nm}$. The much larger value of $\xi_{\rm GL}(0)$ compared to the mean free path ($\ell =0.37$~nm) indicates that the superconductivity in ${\rm Ru_{0.55}Rh_{0.45}P}$ is in dirty-limit.

The BCS coherence length $\xi_{\rm BCS}$  estimated according to  \cite{Tinkham1996}
\be
\xi_{\rm BCS} = \frac{\hbar v_{\rm F}}{\pi \Delta(0)} = \left(\frac{1}{\pi }\right)\frac{\hbar v_{\rm F}}{1.764\,k_{\rm B} T_{\rm c}}
\label{eq:xivF}
\ee
is found to be $\xi_{\rm BCS} = 214$~nm for $v_{\rm F}  = 5.74 \times 10^7~{\rm cm/s}$ and $T_{\rm c} = 3.7$~K\@.  
Within the Ginzburg-Landau thory an estimate of $\lambda_{\rm eff}$ can be obtained using the values of critical fields through the relation \cite{Tinkham1996}
\be
\lambda_{\rm eff}^{2}(0) = \frac{\Phi_0 H_{\rm c2}(0)}{4 \pi H_{\rm c}^2} 
\label{eq:lambda_eff2}
\ee
gives $\lambda_{\rm eff}(0) = 933$~nm. 
The measured and derived superconducting parameters of ${\rm Ru_{0.55}Rh_{0.45}P}$ are listed in Table~\ref{tab:SCParams} together with those of ${\rm Ru_{0.75}Rh_{0.25}As}$.

\subsection{\label{Sec:RuRhP_muSR} Muon spin relaxation and rotation}

\begin{figure}
\includegraphics[width=8cm]{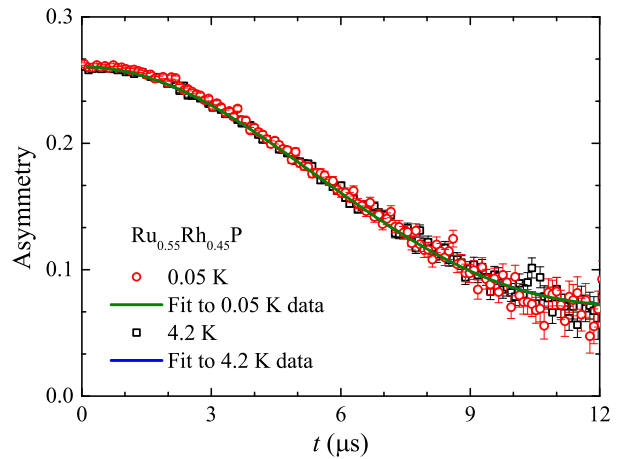} 
\caption{(Color online) Zero field $\mu$SR time spectra for ${\rm Ru_{0.55}Rh_{0.45}P}$ collected at 0.05 and 4.2~K\@. The solid curves show the fits made using Eq.~(\ref{eq:MuSR_ZF}).}
\label{fig:RuRhP_MuSR_ZF}
\end{figure}

The superconducting ground state of ${\rm Ru_{0.55}Rh_{0.45}P}$  was further probed by muon spin relaxation and rotation  measurements. In order to detect magnetic signal associated with the breaking of time-reversal symmetry we first collected the $\mu$SR spectra in zero-field (ZF). The time $t$ evolution of muon spin asymmetry for ZF-$\mu$SR is shown in Fig.~\ref{fig:RuRhP_MuSR_ZF} for 0.05~K and 4.2~K\@. No noticeable change is observed in the muon relaxation rate above (4.2~K $>T_{\rm c}$) and below (0.05~K $<T_{\rm c}$) the superconducting transition temperature which suggests that  the muons do not sense any spontaneous internal field while entering the superconducting state. This indicates that the time-reversal symmetry in the superconducting state is preserved in ${\rm Ru_{0.55}Rh_{0.45}P}$. 

The ZF $\mu$SR spectra are well described by the damped Gaussian Kubo-Toyabe function,
\begin{equation}
\label{eq:MuSR_ZF}
 A_{\rm ZF}(t)=A_0\, G_{\rm KT}(t) \,{\rm e}^{-\Lambda t} + A_{\rm BG},
\end{equation}
where
\begin{equation}
\label{eq:KT}
 G_{\rm KT}(t)=\left[\frac{1}{3}+\frac{2}{3}\left(1-\sigma^2 t^2 \right){\rm e}^{ -\sigma ^2 t^2/2}\right]
\end{equation}
being the Gaussian Kubo-Toyabe function \cite{Hayano1979}, $A_0$ is  the initial asymmetry, $\Lambda$ is the electronic relaxation rate, $\sigma$ is the static relaxation rate, and $A_{\rm BG}$ is the time-independent background contribution. $\sigma$ is a measure of the Gaussian distribution of static fields associated with the nuclear moments and $\Lambda$ accounts for the fluctuating field.  The fits of $\mu$SR spectra by the decay function in Eq.~(\ref{eq:MuSR_ZF}) are shown by solid lines in Fig.~\ref{fig:RuRhP_MuSR_ZF}. The fit yields $\sigma = 0.136(2)~\mu $s$^{-1}$ and $\Lambda = 0.001(1)~\mu$s$^{-1}$ at 0.05~K and $\sigma = 0.136(2)~\mu $s$^{-1}$ and $\Lambda = 0.001(1)~\mu $s$^{-1}$ at 4.2~K\@.  Within the error bar the values of $\sigma$  and $\Lambda$ are essentially the same, indicating that the time reversal symmetry remains preserved..

\begin{figure}
\includegraphics[width=\columnwidth]{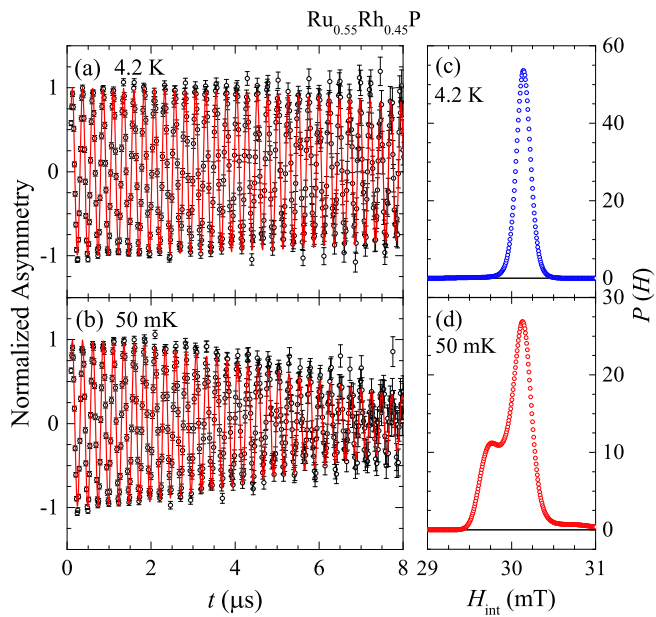} 
\caption{(Color online) Transverse field $\mu$SR time spectra for ${\rm Ru_{0.55}Rh_{0.45}P}$ collected at (a) 4.2~K and (b) 50~mK in an applied magnetic field of 30~mT in the field-cooled state. The solid curves represent the fits made using Eq.~(\ref{eq:MuSR_TF}). The corresponding maximum entropy spectra  are shown in (c) and (d).}
\label{fig:RuRhP_MuSR_TF}
\end{figure}

In order to obtain information about the superconducting gap structure and pairing symmetry we also collected the $\mu$SR spectra in an applied transverse-field (TF).  The TF muon spin precession signals were collected in field-cooled mode with an applied field of 30~mT at 4.2~K (above $T_{\rm c}$) and then the sample was cooled to 0.05~K (below $T_{\rm c}$). The TF-$\mu$SR data were collected at various temperatures in the heating cycle. The TF-$\mu$SR precession signals at 4.2 and 0.05~K are shown in Figs.~\ref{fig:RuRhP_MuSR_TF}(a) and \ref{fig:RuRhP_MuSR_TF}(b). The TF $\mu$SR spectra are well described by an oscillatory function damped with a Gaussian relaxation and an oscillatory background, i.e. by
\begin{equation}
\label{eq:MuSR_TF}
\begin{split}
 A_{\rm TF} (t) &= A_1 \cos\left(\omega_1 t + \phi \right) {\rm e}^{ -\sigma_{\rm TF}^2 t^2/2} \\
  & \hspace{2cm} + A_{\rm BG} \cos\left(\omega_{\rm BG} t + \phi \right)
\end{split}
\end{equation}
where $A_1$ and $A_{\rm BG}$ are the initial asymmetries of sample and background (silver holder), respectively, and $\omega_{1} = \gamma_{\mu}H_{\rm int,1}$ and $\omega_{BG} = \gamma_{\mu}H_{\rm int,BG}$ are the associated muon precession frequencies (with internal field at muon site $H_{\rm int}$ and muon gyromagnetic ratio $\gamma_\mu$); $\phi$ is the initial phase of the muon precession signal. The Gaussian relaxation parameter $\sigma_{\rm TF}$
consists of two contributions: one due to the inhomogeneous field variation across the superconducting vortex lattice $\sigma_{\rm sc}$, and the other due to the nuclear dipolar moments $\sigma_{\rm nm}$ which is assumed to be constant over the entire temperature range. $\sigma_{\rm TF}$ is related to  $\sigma_{\rm sc}$ and $\sigma_{\rm nm}$ as
\begin{equation}
\label{eq:sigma}
 \sigma_{\rm TF}^2 = \sigma_{\rm sc}^2+\sigma_{\rm nm}^2. 
\end{equation}
The nuclear dipolar relaxation rate was obtained by fitting the spectra at $T>T_{\rm c}$, which was then subtracted from $\sigma_{\rm TF}$ according to Eq.~(\ref{eq:sigma}) to obtain the superconducting contribution $\sigma_{\rm sc}$. The fits of the TF $\mu$SR spectra by the decay function in Eq.~(\ref{eq:MuSR_TF}) are shown by solid red curves in Figs.~\ref{fig:RuRhP_MuSR_TF}(a) and \ref{fig:RuRhP_MuSR_TF}(b). At low tempearture, e.g. at $T =0.05$~K ($T<T_{\rm c}$), the $\sigma_{\rm TF}$ is found to be much larger than that at $T>T_{\rm c}$. Such an increase of $\sigma_{\rm TF}$ is due to the vortex lattice formation and reveals bulk superconductivity in ${\rm Ru_{0.55}Rh_{0.45}P}$.

The maximum entropy spectra that depict the magnetic field probability distribution $P(H)$ corresponding to the TF $\mu$SR spectra at 4.2~K and 0.05~K in Figs.~\ref{fig:RuRhP_MuSR_TF}(a) and \ref{fig:RuRhP_MuSR_TF}(b) are shown in Figs.~\ref{fig:RuRhP_MuSR_TF}(c) and \ref{fig:RuRhP_MuSR_TF}(d), respectively. It is seen from Figs.~\ref{fig:RuRhP_MuSR_TF}(c) and \ref{fig:RuRhP_MuSR_TF}(d)  that in the normal state (at 4.2~K) a sharp peak is observed at $H_{\rm int}$ centered around the applied $H$, whereas in superconducting state (at 0.05~K) an additional broad peak appears at a lower field ($H_{\rm int} < H$). The appearance of an additional peak at an internal field lower than the applied $H$ is a characteristic of a type-II behavior (due to the field distribution of the flux-line lattice in the vortex state) and indicates a type-II superconductivity in ${\rm Ru_{0.55}Rh_{0.45}P}$ as also inferred from the bulk properties measurements and $\kappa_{\rm GL}$ listed in Table~\ref{tab:SCParams}. 

\begin{figure}
\includegraphics[width=8cm]{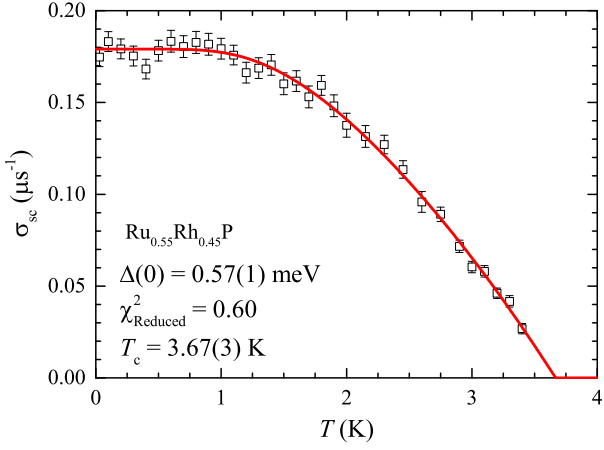} 
\caption{(Color online) Temperature $T$ dependence of the muon spin relaxation rate $\sigma_{\rm sc}$ for ${\rm Ru_{0.55}Rh_{0.45}P}$ collected in an applied transverse field of 30~mT in field cooled state. The solid curve represents the fit for an isotropic single gap $s$-wave model according to Eq.~(\ref{eq:sigma_TF}).}
\label{fig:RuRhP_MuSR_Sigma}
\end{figure}

The $\sigma_{\rm sc}(T)$ obtained from $\sigma_{\rm TF} (T)$ is shown in Fig.~\ref{fig:RuRhP_MuSR_Sigma}. The $\sigma_{\rm sc}$ is directly related to the magnetic penetration depth and superfluid density and therefore carries information about the symmetry and size of superconducting gap. As the TF spectra were collected at 30~mT which is much smaller than the upper critical field, following Brandt \cite{Brandt2003}, for a triangular vortex lattice $\sigma_{\rm sc}$ is related to the effective penetration depth $\lambda_{\rm eff}$ as
\begin{equation}
\label{eq:sigma_lambda}
\frac{\sigma_{\rm sc}}{\gamma_{\mu}} = \sqrt{0.00371} \frac{\Phi_0}{\lambda_{\rm eff}^2}.
\end{equation}
This relation is valid for $0.13/ \kappa^{2}  {\ll} (H/H_{c2}) {\ll} 1$ and  $\kappa \gg 70$ \cite{Brandt2003} and these conditions are approximated by the parameters listed in Table~\ref{tab:SCParams} for ${\rm Ru_{0.55}Rh_{0.45}P}$. The superconducting gap can be modeled by \cite{Prozorov}
\begin{multline}
\frac{\sigma_{\rm sc}(T)}{\sigma_{\rm sc}(0)}=\frac{\lambda_{\rm eff}^{-2}(T,\Delta)}{\lambda_{\rm eff}^{-2}(0)} \\
= 1 + \frac{1}{\pi} \int_{0}^{2\pi} \int_{\Delta(T, \varphi)}^{\infty}\frac{\partial f}{\partial E}\frac{E\,{\rm d}E\,{\rm d}\varphi}{\sqrt{E^2-\Delta^2(T, \varphi)}},
\label{eq:sigma_TF}
\end{multline}
\noindent where $f=\left[1+\exp\left(-E/k_{\mathrm{B}}T\right)\right]^{-1}$ is the Fermi function and $\varphi$ is the azimuthal angle along the Fermi surface. The $T$ and $\varphi$ dependent order parameter $\Delta(T, \varphi) = \Delta(0) \delta(T/\it {T}_c)g(\varphi)$, where the function $g(\varphi)$ contains the angular dependence of the superconducting gap function. For an isotropic gap $s$-wave model there is no angular dependence and hence we used $g(\varphi)= 1$ \cite{Annett, Pang}. We used the BCS approximation $\delta(T/T_c) =\tanh[(1.82){(1.018(T_c/T-1))}^{0.51}]$~\cite{UBe131}.

The $\sigma_{\rm sc}(T)$ data could be described well by a single-band isotropic gap $s$-wave model according to Eq.~(\ref{eq:sigma_TF}). The fit is shown by the solid red curve in Fig.~\ref{fig:RuRhP_MuSR_Sigma}. The fit yielded $\Delta(0) = 0.57(1)$~meV which in turn gives $\Delta(0)/k_{\rm B}T_{\rm c} = 1.78(3)$ which is in very good agreement with the expected BCS value of 1.764. From the fit of $\sigma_{\rm sc}(T)$ we get $\sigma_{\rm sc}(0) = 0.179(2)~\mu {\rm s}^{-1}$ which according to Eq.~(\ref{eq:sigma_lambda}) yields $\lambda_{\rm eff} = 309(3)$~nm. This observed value of $\lambda_{\rm eff}$ is much lower than the calculated value of $\lambda_{\rm eff} = 933$~nm (see Table~\ref{tab:SCParams}). As the $\mu$SR provides a reliable estimate of superfluid density, the value of $\lambda_{\rm eff}$ obtained through the analysis of $\mu$SR is more realistic. The results discussed above that were obtained from the $\mu$SR data (particularly the temperature dependence of $\sigma_{\rm sc}$, which fits better to a single $s$-wave gap with the BCS expected value of $\Delta(0)/k_{\rm B}T_{\rm c}$) together reflect a single-band fully gapped isotropic $s$-wave singlet pairing weakly coupled conventional type-II superconductivity in ${\rm Ru_{0.55}Rh_{0.45}P}$.
Our $\mu$SR  data thus reflect a single-band fully gapped isotropic $s$-wave singlet pairing weakly coupled conventional type-II superconductivity in ${\rm Ru_{0.55}Rh_{0.45}P}$.

\section{\label{Sec:RuRhAs} Superconductivity in \bm{${\rm R\lowercase{u}_{0.75}R\lowercase{h}_{0.25}A\lowercase{s}}$} }

\subsection{\label{Sec:RuRhAs_MT} Magnetic Susceptibility and Magnetization}

\begin{figure}
\includegraphics[width=8cm]{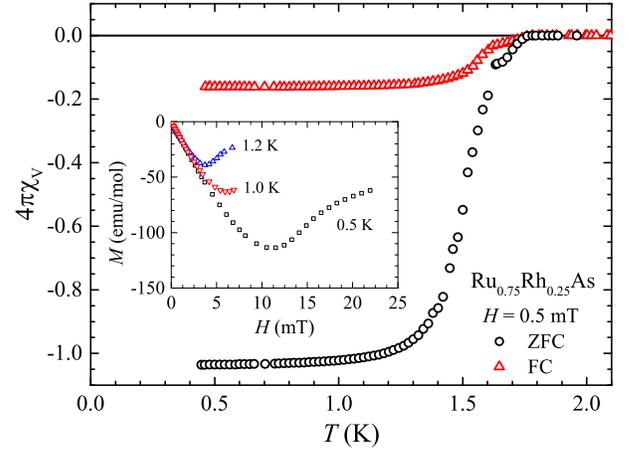} 
\caption{(Color online) Zero field cooled (ZFC) and field cooled (FC) magnetic susceptibility $\chi$ data of  ${\rm Ru_{0.75}Rh_{0.25}As}$ as a function of temperature~$T$ for 0.45~K~$\leq T \leq$~2~K measured in applied magnetic field $H=0.5$~mT. Inset: Isothermal magnetization $M(H)$ data measured at indicated temperatures.}
\label{fig:RuRhAs_chi}
\end{figure}

The ZFC and FC $\chi(T)$ data for ${\rm Ru_{0.75}Rh_{0.25}As}$ measured in $H=0.5$~mT are shown in Fig.~\ref{fig:RuRhAs_chi}. Both ZFC and FC $\chi(T)$ show clear superconducting transition, an onset of superconductivity is seen at 1.73~K followed by a sharp transition below 1.63~K\@. Further, the large Meissner signal for the ZFC $\chi$ reveals bulk superconductivity with a superconducting phase fraction of $\sim 100$\%. The isothermal $M(H)$ data also show a large Meissner signal (inset of Fig.~\ref{fig:RuRhAs_chi}). At $T=0.5$~K, $M(H)$ is linear for fields $\sim 4$~mT and deviates thereafter. This linear regime and hence $H_{\rm c1}$ decreases with increasing $T$ as the temperature approaches $T_{\rm c}$. The $T$ dependence of $H_{\rm c1}$ inferred from the $M(H)$ isotherms is discussed latter.

\subsection{\label{Sec:RuRhAs_Rho} Electrical Resistivity}

\begin{figure}
\includegraphics[width=8cm]{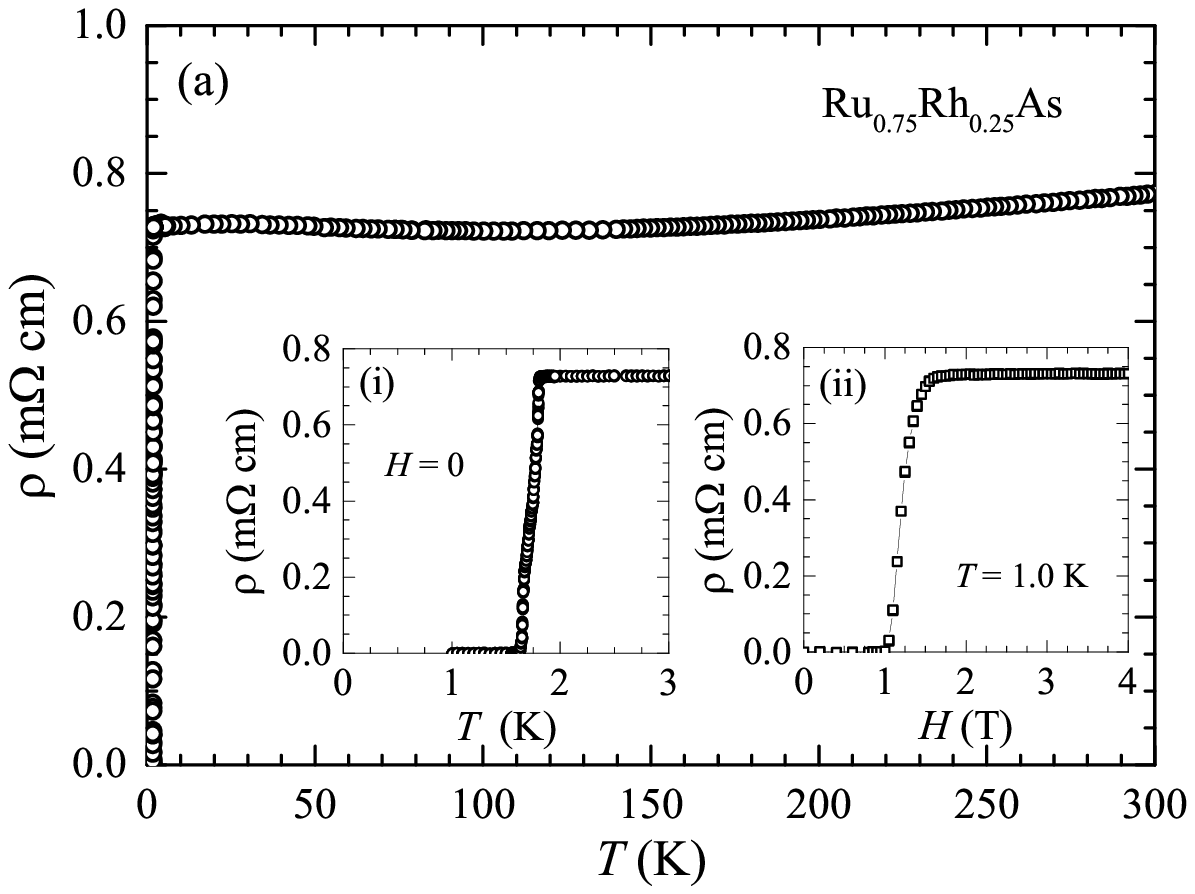}\vspace{0.1cm}
\includegraphics[width=8cm]{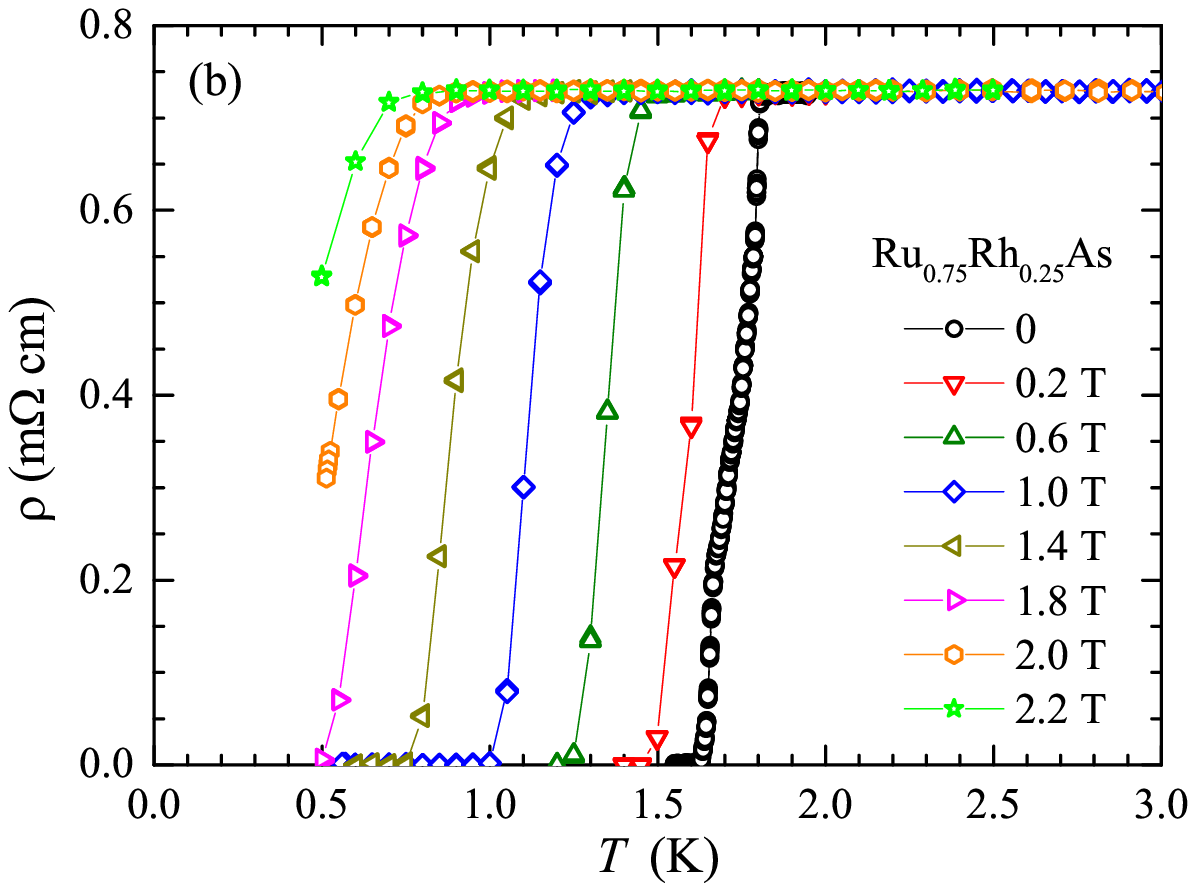}
\caption{(Color online) (a) Electrical resistivity $\rho$ of ${\rm Ru_{0.75}Rh_{0.25}As}$ as a function of temperature~$T$ for 0.45~K~$\leq T \leq$~300~K measured in applied magnetic field $H=0$. Insets: (i) Expanded plot of $\rho(T)$ showing the superconducting transition, and  (ii) $\rho(H)$ at 1~K\@.
(b) Low-$T$ $\rho(T)$ at different $H$ for 0.45~K~$\leq T \leq$~3~K.}
\label{fig:RuRhAs_rho}
\end{figure}

The $\rho(T)$ data of ${\rm Ru_{0.75}Rh_{0.25}As}$ measured with various applied fields are shown in Fig.~\ref{fig:RuRhAs_rho}. The $\rho$ exhibits metallic behavior and undegoes a superconducting transition.  The residual resistivity just before entering the superconducting state is 7.25~m$\Omega$\,cm  and the residual resistivity ratio is $\sim 1.1$. The  $T_{\rm c}^{\rm  onset}$ for superconductivity is $ \approx 1.80$~K with the zero resistance state below $T_{\rm c\, 0} \approx 1.63$~K\@ [inset (i) of Fig.~\ref{fig:RuRhAs_rho}(a)]. The  $\rho(T)$ measured in different $H$ shown in Fig.~\ref{fig:RuRhAs_rho}(b) shows the suppression of $T_{\rm c}$ by field, $T_{\rm c}$ decreases with increasing $H$. The $\rho(H)$ data in inset (ii) of Fig.~\ref{fig:RuRhAs_rho}(a) indicate that a field of $\sim 1.6$~T would be required to destroy the superconductivity in ${\rm Ru_{0.75}Rh_{0.25}As}$.

\subsection{\label{Sec:RuRhAs_HC} Heat Capacity}

\begin{figure}
\includegraphics[width=8cm]{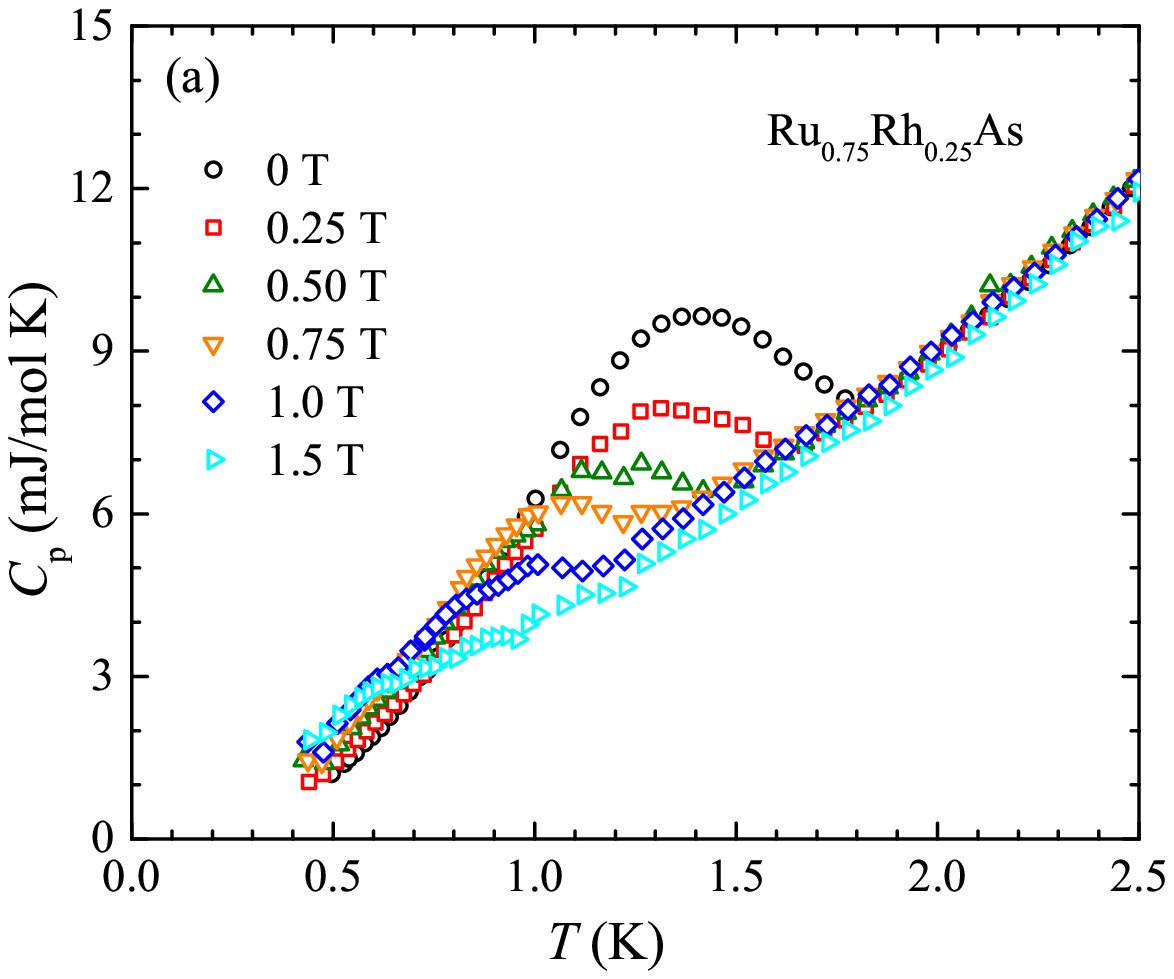}\vspace{0.1in}
\includegraphics[width=8cm]{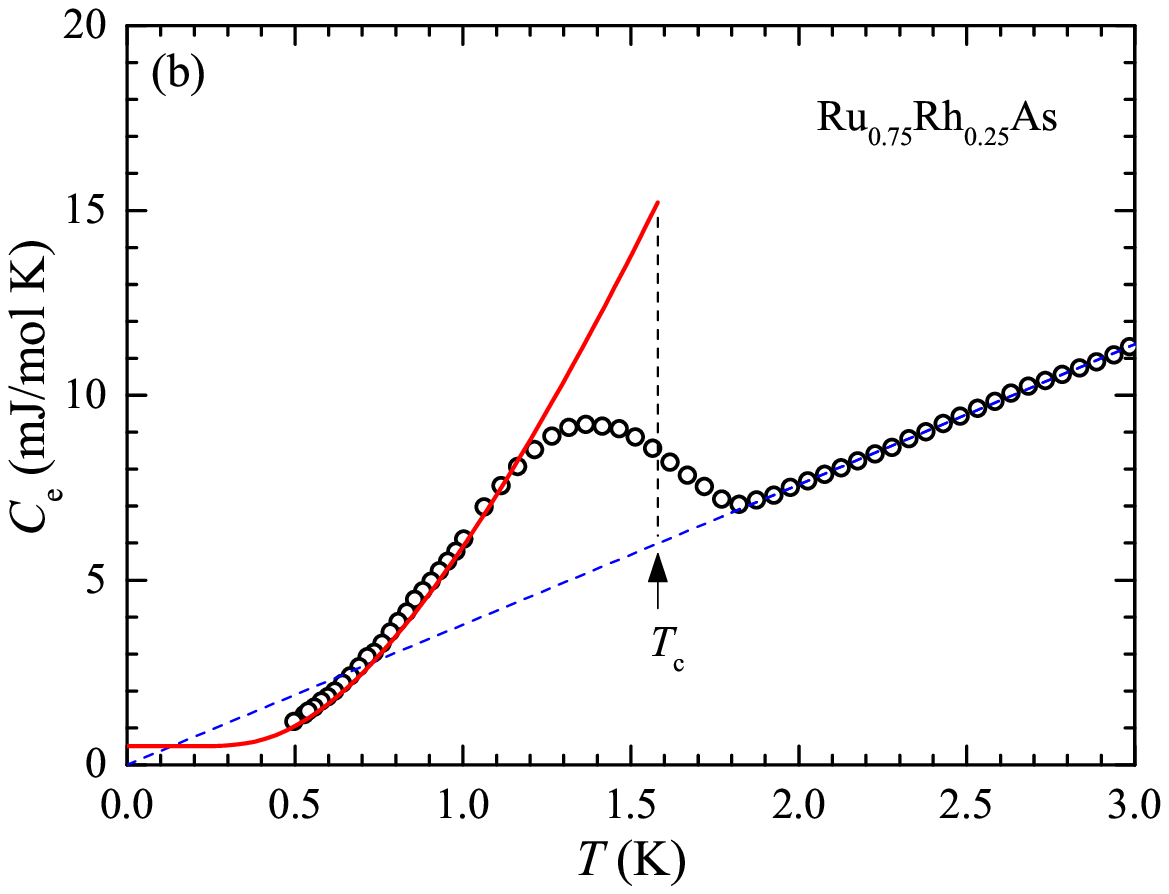}
\caption{(Color online) (a) Heat capacity $C_{\rm p}$ of ${\rm Ru_{0.75}Rh_{0.25}As}$ as a function of temperature $T$ for 0.45~K~$\leq T \leq$~2.5~K measured in different indicated applied magnetic fields. (b) Electronic contribution $C_{\rm e}$ to zero field heat capacity as a function of temperature $T$. The solid red curve is the theoretical prediction for single-band fully gapped [$\Delta(0)/k_{\rm B}T_{\rm c} = 1.764$] BCS superconductivity. The theoretical curve is shifted up by 0.50~mJ/mol~K which accounts for nonsuperconducting contribution to $C_{\rm e}$. }
\label{fig:RuRhAs_HC}
\end{figure}

The $C_{\rm p}(T)$ data of ${\rm Ru_{0.75}Rh_{0.25}As}$  measured with various applied fields  are shown in Fig.~\ref{fig:RuRhAs_HC}(a). The $C_{\rm p}(T)$ shows a clear anomaly related to the superconducting transition. An onset of superconductivity is seen at $T_{\rm c}^{\rm  onset} = 1.77$~K in zero field $C_{\rm p}(T)$ data. A $T_{\rm c} = 1.60(4)$~K\@ is obtained by the entropy-conserving construction shown in Fig.~\ref{fig:RuRhAs_HC}(b). As expected, the application of magnetic field suppresses the $T_{\rm c}$. In addition, the field also broadens the peak. 

\begin{figure*}
\includegraphics[width=\textwidth, keepaspectratio]{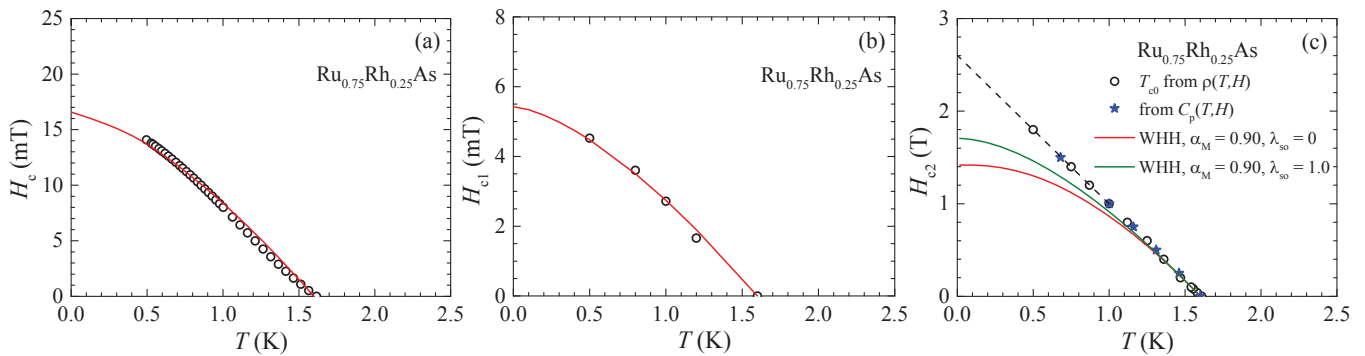}
\caption{\label{fig:LaIrSi3_Hcritical} (Color online) (a) Thermodynamic critical field $H_{c}$ of ${\rm Ru_{0.75}Rh_{0.25}As}$ as a function of temperature $T$  obtained from the experimental electronic heat capacity $C_{\rm e}(T)$ data. (b) Lower critical field $H_{c1}(T)$ obtained from $M(H)$ data, and (c) Upper critical field $H_{c2}(T)$ obtained from $C_{\rm p}(T,H)$ and $\rho(T,H)$ data. The solid curves represent the fits as discussed in text. The dashed line in (c) shows a linear behavior.}
\label{fig:RuRhAs_HCrit}
\end{figure*}

From the analysis of normal state low-$T$ $C_{\rm p}(T)$ data we obtain $ \gamma_n = 3.79(6)$~mJ/mol\,K$^2$ and $\beta = 0.169(2)$~mJ/mol\,K$^4$.  The density of states at the Fermi level is estimated to be ${\cal D}(E_{\rm F}) = 1.61(2) $ states/eV f.u. for both spin directions. The Debye temperature is found to be $\Theta_{\rm D} = 284(2)$~K \cite{Kittel2005}. The electron-phonon coupling constant estimated according to Eq.~(\ref{eq:lambda}) for $T_{\rm c} = 1.6$~K and $\Theta_{\rm D}= 284$~K is $\lambda_{\rm {e-ph}} = 0.49$ which reflects a weak-coupling superconductivity in ${\rm Ru_{0.75}Rh_{0.25}As}$. For $\lambda_{\rm {e-ph}} = 0.49$ the bare band-structure density of states is found to be ${\cal D}_{\rm band}(E_{\rm F}) = 1.08$ states/eV f.u. for both spin directions, and the effective quasiparticle mass turns out to be $m^\ast =1.49 \,m_{\rm e}$. The Fermi velocity and mean free path are found to be $v_{\rm F} = 2.08 \times 10^8~{\rm cm/s}$ and $\ell = 0.023$~nm. We note that the estimated value of $\ell$ is significantly lower than the lattice constant suggesting that the Drude model of electrical conduction fails to account for the measured resitivity.

\subsection{\label{Sec:RuRhAs_SC_Prop} Superconducting state properties}

The  electronic contribution $C_{\rm e}(T)$ to heat capacity of ${\rm Ru_{0.75}Rh_{0.25}As}$ is shown in Fig.~\ref{fig:RuRhAs_HC}(b) which clearly shows the bulk nature of superconductivity.  Utilizing the entropy-conserving construction in Fig.~\ref{fig:RuRhAs_HC}(b) we obtain $\Delta C_{\rm e} = 8.14(8)$~mJ/mol\,K at $T_{\rm c}$ and $\Delta C_{\rm e}/ \gamma_{\rm n} T_{\rm c} = 1.42(2)$ for $T_{\rm c} = 1.6$~K and  $\gamma_{\rm n} = 3.79$~mJ/mol\,K$^2$ in very good agreement with the weak-coupling BCS value of 1.426.  The theoretical prediction for a single-band fully gapped BCS superconductor is shown in Fig.~\ref{fig:RuRhAs_HC}(b) and there is very reasonable agreement with the experimental data. The theoretical curve is shifted up by 0.50~mJ/mol~K to account for the presence of small nonsuperconducting impurity phase(s) in sample.

The thermodynamic critical field estimated from the zero-field heat capacity data is shown in Fig.~\ref{fig:RuRhAs_HCrit}(a). The $H_{\rm c}(T)$ data follow the behavior $H_{\rm c}(T) = H_{\rm c}(0) [1-(T/T_{\rm c})^p]$, with $p = 1.5$. The fit of $H_{\rm c}(T)$ data by this behavior is shown by the solid red curve in Fig.~\ref{fig:RuRhP_HCrit}(a), giving $H_{\rm c}(0) = 16.6(2)$~mT. 

The lower critical field determined from the $M(H)$ data is shown in Fig.~\ref{fig:RuRhAs_HCrit}(b) as a function of temperature. The $H_{\rm c1}(T)$ data follow $H_{\rm c1}(T) = H_{\rm c1}(0) [1-(T/T_{\rm c})^p]$, with $p = 1.5$. The fit of $H_{\rm c1}(T)$ by this expression is shown by the solid red curve in Fig.~\ref{fig:RuRhP_HCrit}(b) which gives a $H_{\rm c1}(0) = 5.4(1)$~mT. Similar to the case of ${\rm Ru_{0.55}Rh_{0.45}P}$, the small value of $H_{\rm c1}(0)$ compared to the value of $H_{\rm c}(0)$ indicates a type-II superconductivity in ${\rm Ru_{0.75}Rh_{0.25}As}$.

The temperature dependence of upper critical field determined from the $C_{\rm p}(T,H)$ and $\rho(T,H)$ data is shown in Fig.~\ref{fig:RuRhAs_HCrit}(c). With an initial slope of $dH_{\rm c2}(T)/dT|_{T = T\rm c} = -1.72(4)$~T/K, the orbital critical field  $H_{\rm c2}^{\rm Orb} (0) = 2.01(4)$~T in the clean limit and $H_{\rm c2}^{\rm Orb} (0) = 1.90(4)$~T in the dirty limit. The Pauli-limiting upper critical field is found to be $H_{\rm P}(0) = 2.98(7)$~T\@, accordingly we obtain Maki parameter $\alpha_{\rm M}= 0.90$. The $\alpha_{\rm M}$ is close to 1 and suggests that the Pauli limiting is playing role in determining the $H_{\rm c2}$. Similar to the case of ${\rm Ru_{0.55}Rh_{0.45}P}$, the $H_{\rm c2}(T)$ of ${\rm Ru_{0.75}Rh_{0.25}As}$ shows a linear behavior that cannot be described by the WHH model. The WHH model predictions for $\alpha_{\rm M} = 0.90$ and $\lambda_{\rm so} = 0$ as well as $\lambda_{\rm so} = 1.0$ are shown in Fig.~\ref{fig:RuRhAs_HCrit}(c) to show the departure from the WHH model, particularly at low-$T$. A linear extrapolation of $H_{\rm c2}(T)$ yields $H_{\rm c2}(0) = 2.60(1)$~T\@.

The Ginzburg-Landau parameter estimated from $H_{\rm c2}(0) = 2.60$~T and $H_{\rm c}(0) = 16.6$~mT   is $\kappa_{\rm GL} \approx 111$, characterizing ${\rm Ru_{0.75}Rh_{0.25}As}$ as a type-II superconductor.  The Ginzburg-Landau coherence length is found to be $\xi_{\rm GL}(0) = 11~{\rm nm}$. The $\xi_{\rm GL}(0)$ is very large compared to the mean free path ($\ell = 0.023$~nm), suggesting a dirty-limit superconductivity in ${\rm Ru_{0.75}Rh_{0.25}As}$. For $v_{\rm F}  = 2.08 \times 10^8~{\rm cm/s}$ and $T_{\rm c} = 1.6$~K, the BCS coherence length is found to be $\xi_{\rm BCS} = 1792$~nm.  The effective magnetic penetration depth  is estimated to be $\lambda_{\rm eff}(0) = 1247$~nm. The measured and derived superconducting parameters of ${\rm Ru_{0.75}Rh_{0.25}As}$ are listed in Table~\ref{tab:SCParams}. 

\subsection{\label{Sec:RuRhAs_muSR} Muon spin relaxation and rotation}

\begin{figure}
\includegraphics[width=8cm]{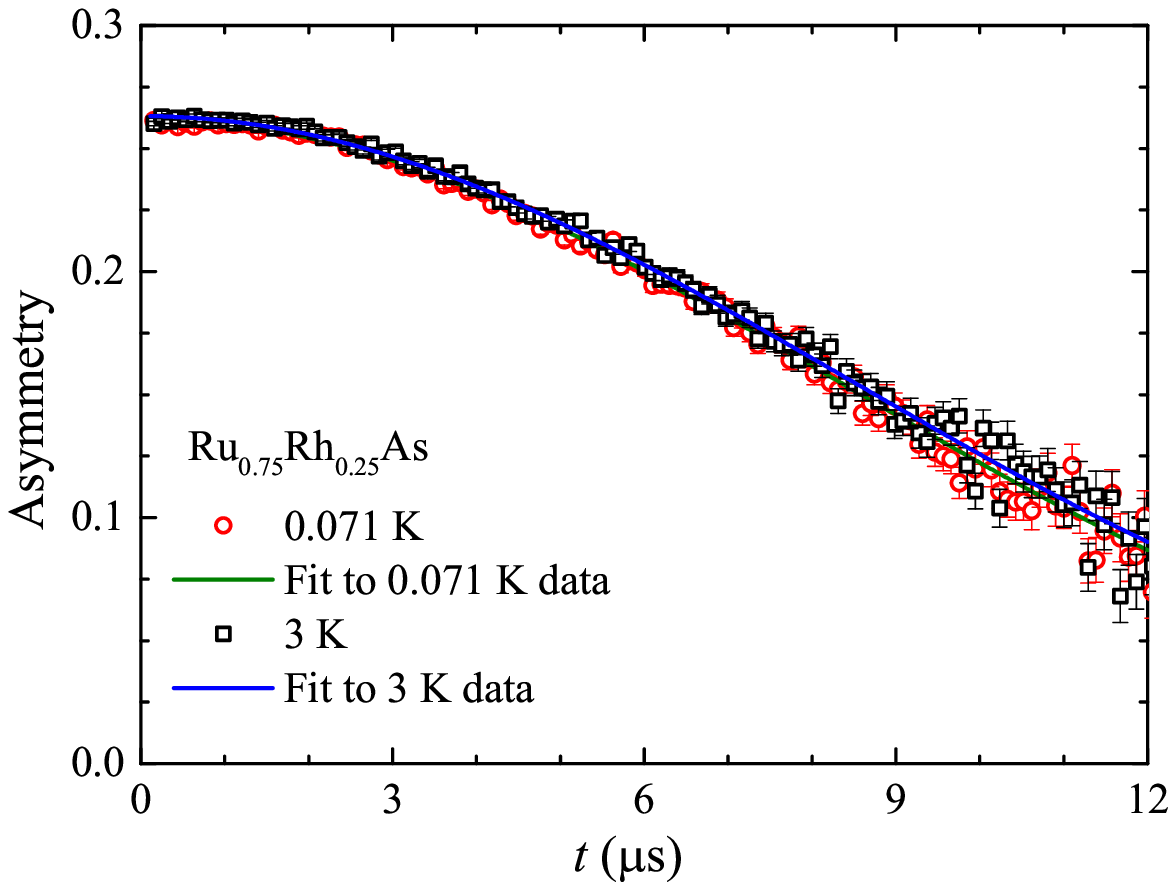} 
\caption{(Color online) Zero field $\mu$SR time spectra for ${\rm Ru_{0.75}Rh_{0.25}As}$ collected at 0.071 and 3~K\@. The solid curves show the fits made using Eq.~(\ref{eq:MuSR_ZF}).}
\label{fig:RuRhAs_MuSR_ZF}
\end{figure}

In order to further probe the superconducting ground state of ${\rm Ru_{0.75}Rh_{0.25}As}$ we also carried muon spin relaxation and rotation measurements both in zero field and transverse field. The ZF-$\mu$SR spectra are shown in Fig.~\ref{fig:RuRhAs_MuSR_ZF} for 0.071 and 3~K\@. As seen from Fig.~\ref{fig:RuRhAs_MuSR_ZF} the muon relaxation rate above (3~K) and below (0.071~K) $T_{\rm c}$ are very similar which indicates that the time-reversal symmetry is preserved in the superconducting state of ${\rm Ru_{0.75}Rh_{0.25}As}$. The ZF $\mu$SR spectra were analyzed by damped Gaussian Kubo-Toyabe function given in Eq.~(\ref{eq:MuSR_ZF}), the fits of $\mu$SR spectra are shown by solid lines in Fig.~\ref{fig:RuRhAs_MuSR_ZF}. From the fits we obtained $\sigma = 0.089(1)~\mu $s$^{-1}$ and $\Lambda = 0(0)~\mu$s$^{-1}$ at 0.071~K and $\sigma = 0.088(2)~\mu $s$^{-1}$ and $\Lambda = 0(0)~\mu $s$^{-1}$ at 3~K\@.  

\begin{figure}
\includegraphics[width=\columnwidth]{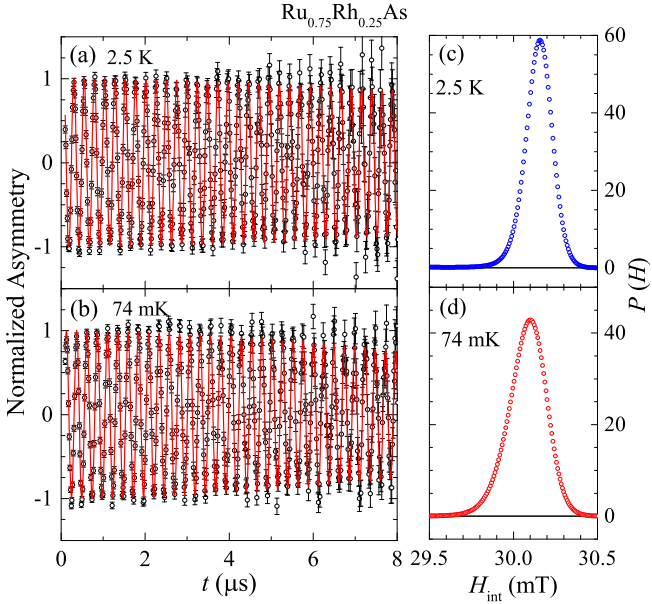} 
\caption{(Color online) Transverse field $\mu$SR time spectra for ${\rm Ru_{0.75}Rh_{0.25}As}$ collected at (a) 2.5~K and (b) 74~mK\@ in an applied magnetic field of 30~mT in the field-cooled state. The solid curves represent the fits made using Eq.~(\ref{eq:MuSR_TF}). The corresponding maximum entropy spectra  are shown in (c) and (d).}
\label{fig:RuRhAs_MuSR_TF}
\end{figure}

The TF $\mu$SR spectra of ${\rm Ru_{0.75}Rh_{0.25}As}$, which were collected in field-cooled mode with an applied field of 30~mT, at 2.5~K (above $T_{\rm c}$) and 0.074~K (below $T_{\rm c}$) are shown in Figs.~\ref{fig:RuRhAs_MuSR_TF}(a) and \ref{fig:RuRhAs_MuSR_TF}(b). The TF $\mu$SR spectra were analyzed by an oscillatory function damped with a Gaussian combined with an oscillatory background given in Eq.~(\ref{eq:MuSR_TF}). The fits of the TF $\mu$SR spectra are shown by solid red curves in Figs.~\ref{fig:RuRhAs_MuSR_TF}(a) and \ref{fig:RuRhAs_MuSR_TF}(b). The $\sigma_{\rm TF}$ is found to be significantly larger at $T<T_{\rm c}$ (e.g. at $T =0.074$~K) compared to that at $T>T_{\rm c}$, thus revealing a bulk superconductivity in ${\rm Ru_{0.75}Rh_{0.25}As}$. The maximum entropy spectra corresponding to the TF $\mu$SR spectra at 2.5 and 0.074~K are shown in Figs.~\ref{fig:RuRhAs_MuSR_TF}(c) and \ref{fig:RuRhAs_MuSR_TF}(d), respectively. Only one peak (centered around the applied $H$) is observed in both normal state (at 2.5~K) and superconducting state (at 0.074~K), however, at 0.074~K the peak broadens a little with an extra shoulder on lower field side indicating type-II superconductivity. This observation for ${\rm Ru_{0.75}Rh_{0.25}As}$ is different from that in ${\rm Ru_{0.55}Rh_{0.45}P}$  where an additional peak at an internal field lower than the applied $H$ was clearly observed.

\begin{figure}
\includegraphics[width=8cm]{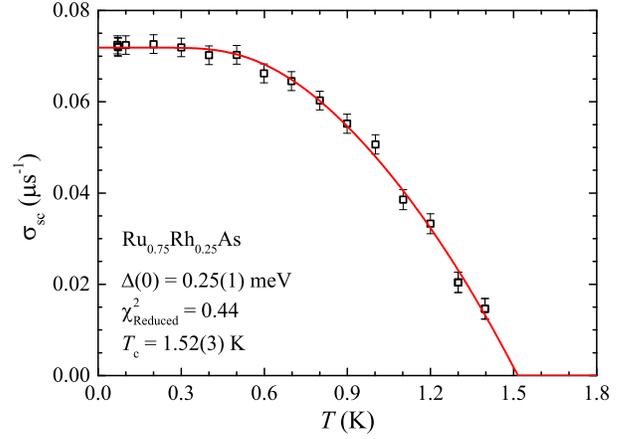} 
\caption{(Color online) Temperature $T$ dependence of the muon spin relaxation rate $\sigma_{\rm sc}$ for ${\rm Ru_{0.75}Rh_{0.25}As}$ collected in an applied transverse field of 30~mT in the field-cooled state. The solid curve represents the fit for an isotropic single gap $s$-wave model according to Eq.~(\ref{eq:sigma_TF}).}
\label{fig:RuRhAs_MuSR_Sigma}
\end{figure}

The $\sigma_{\rm sc}(T)$ obtained according to Eq.~(\ref{eq:sigma}) from $\sigma_{\rm TF} (T)$ of ${\rm Ru_{0.75}Rh_{0.25}As}$ is shown in Fig.~\ref{fig:RuRhAs_MuSR_Sigma}. The condition $0.13/ \kappa^{2}  {\ll} (H/H_{c2}) {\ll} 1$ and  $\kappa \gg 70$ \cite{Brandt2003} are fulfilled by the parameters of ${\rm Ru_{0.75}Rh_{0.25}As}$ listed in Table~\ref{tab:SCParams}, therefore  $\sigma_{\rm sc}$ can be related to the effective penetration depth $\lambda_{\rm eff}$ according to Eq.~(\ref{eq:sigma_lambda}) and the superconducting gap can be modeled by Eq.~(\ref{eq:sigma_TF}) similar to the case of ${\rm Ru_{0.55}Rh_{0.45}P}$ discussed above. Similar to ${\rm Ru_{0.55}Rh_{0.45}P}$ the $\sigma_{\rm sc}(T)$ of ${\rm Ru_{0.75}Rh_{0.25}As}$ is also very well described by the single band isotropic gap $s$-wave model. The fit of $\sigma_{\rm sc}(T)$ by Eq.~(\ref{eq:sigma_TF}) is shown by the solid red curve in Fig.~\ref{fig:RuRhAs_MuSR_Sigma}. The fit yielded $\Delta(0) = 0.25(1)$~meV which corresponds to $\Delta(0)/k_{\rm B}T_{\rm c} = 1.81(6)$ which within the error bar is in very good agreement with the expected BCS value of 1.764. From the value of $\sigma_{\rm sc}(0) = 0.072(1)~\mu {\rm s}^{-1}$  we obtain $\lambda_{\rm eff} = 487(4)$~nm which is again substantially lower than the calculated value (see Table~\ref{tab:SCParams}). Similar to the case of ${\rm Ru_{0.55}Rh_{0.45}P}$, the $\mu$SR data of ${\rm Ru_{0.75}Rh_{0.25}As}$ also reflect a weakly coupled single-band fully gapped isotropic $s$-wave singlet pairing conventional type-II superconductivity.

\section{\label{Conclusion} Conclusions}

We have investigated the superconductiing properties of two pseudo-binary pnictides ${\rm Ru_{0.55}Rh_{0.45}P}$ and ${\rm Ru_{0.75}Rh_{0.25}As}$ through $ \chi(T)$, $M(H)$, $C_{\rm p}(T, H)$, $\rho(T,H)$ and $\mu$SR measurements. The $\chi(T)$, $C_{\rm p}(T)$ and $\rho(T)$ present conclusive evidence for bulk superconductivity below 3.7~K in ${\rm Ru_{0.55}Rh_{0.45}P}$ and  below 1.6~K in ${\rm Ru_{0.75}Rh_{0.25}As}$. The superconducting state electronic heat capacity of both ${\rm Ru_{0.55}Rh_{0.45}P}$ and ${\rm Ru_{0.75}Rh_{0.25}As}$ follows BCS superconductivity characterized by $\Delta C_{\rm e}/\gamma_{\rm n} T_{\rm c} = 1.426$ and  $\Delta(0)/k_{\rm B}T_{\rm c} = 1.764$. Various normal and superconducting state parameters have been estimated and a weakly-coupled electron-phonon driven type-II superconductivity in dirty-limit is inferred for both ${\rm Ru_{0.55}Rh_{0.45}P}$ and ${\rm Ru_{0.75}Rh_{0.25}As}$. 

For both ${\rm Ru_{0.55}Rh_{0.45}P}$ and ${\rm Ru_{0.75}Rh_{0.25}As}$, the upper critical field is found to exhibit a linear temperature dependence, which could not be described by the isotropic dirty limit theory of WHH. This type of linear behavior has been associated with two band superconductivity, however, our $\mu$SR  data do not support two band superconductivity in these compounds. The $\mu$SR data confirm the conventional type-II behavior and reveal that the time reversal symmetry is preserved in both the compounds. The analysis of the temperature dependence of the superconducting contribution to muon relaxation rate $\sigma_{\rm sc}(T)$ obtained from the TF-$\mu$SR data reveals an isotropic single gap $s$-wave superconductivity in both the compounds. 

\acknowledgments
VKA and BL acknowledge Helmholtz Gemeinschaft for funding via the Helmholtz Virtual Institute (Project No. VH-VI-521). We would like to thank the ISIS facility for providing beam time on the MuSR spectrometer, RB1710170.

\end{document}